\PassOptionsToPackage{english}{babel}
\documentclass[reprint, superscriptaddress, amsmath, amssymb, aps, prd, nofootinbib]{revtex4-1}
\usepackage[english]{babel}
\usepackage{graphicx}
\usepackage{bm}
\usepackage{hyperref}
\hypersetup{
    colorlinks=true,
    linkcolor=blue,
    filecolor=magenta,      
    urlcolor=blue,
    citecolor=blue
    }
\usepackage{geometry}
\usepackage{bm}
\usepackage{braket}
\usepackage{listings}
\usepackage{etoolbox}
\usepackage{amsfonts}
\usepackage{amsmath}
\newcommand{\be}{\begin{equation}}
\newcommand{\ee}{\end{equation}}
\renewcommand{\vec}[1]{\mathbf{#1}}
\newcommand{\Ai}{\operatorname{Ai}}
\newcommand{\Bi}{\operatorname{Bi}}

\begin{document}

\title{An efficient method for solving highly oscillatory ordinary differential equations with applications to physical systems}
\author{F.\ J.\ Agocs}
\email{fa325@cam.ac.uk}
\affiliation{Astrophysics Group, Cavendish Laboratory, J.\ J.\ Thomson Avenue, Cambridge, CB3 0HE, UK}
\affiliation{Kavli Institute for Cosmology, Madingley Road, Cambridge, CB3 0HA, UK}
\author{W.\ J.\ Handley}
\email{wh260@cam.ac.uk}
\affiliation{Astrophysics Group, Cavendish Laboratory, J.\ J.\ Thomson Avenue, Cambridge, CB3 0HE, UK}
\affiliation{Kavli Institute for Cosmology, Madingley Road, Cambridge, CB3 0HA, UK}
\author{A.\ N.\ Lasenby}
\email{a.n.lasenby@mrao.cam.ac.uk}
\affiliation{Astrophysics Group, Cavendish Laboratory, J.\ J.\ Thomson Avenue, Cambridge, CB3 0HE, UK}
\affiliation{Kavli Institute for Cosmology, Madingley Road, Cambridge, CB3 0HA, UK}
\author{M.\ P.\ Hobson}
\email{mph@mrao.cam.ac.uk}
\affiliation{Astrophysics Group, Cavendish Laboratory, J.\ J.\ Thomson Avenue, Cambridge, CB3 0HE, UK}
\date{\today}

\begin{abstract}

    We present a novel numerical routine (\texttt{oscode}) with a C++ and Python interface for the efficient solution of one-dimensional, second-order, ordinary differential equations with rapidly oscillating solutions.
    The method is based on a Runge--Kutta-like stepping procedure that makes use of the Wentzel--Kramers--Brillouin (WKB) approximation to skip regions of integration where the characteristic frequency varies slowly.
    In regions where this is not the case, the method is able to switch to a made-to-measure Runge--Kutta integrator that minimises the total number of function evaluations.
    We demonstrate the effectiveness of the method with example solutions of the Airy equation and an equation exhibiting a burst of oscillations, discussing the error properties of the method in detail.
    We then show the method applied to physical systems. First, the one-dimensional, time-independent Schr\"odinger equation is solved as part of a shooting method to search for the energy eigenvalues for a potential with quartic anharmonicity.
    Then, the method is used to solve the Mukhanov--Sasaki equation describing the evolution of cosmological perturbations, and the primordial power spectrum of the perturbations is computed in different cosmological scenarios.
    We compare the performance of our solver in calculating a primordial power spectrum of scalar perturbations to that of \texttt{BINGO}, an efficient code specifically designed for such applications, and find that our method performs better.
    
\end{abstract}

\maketitle

\section{Introduction}
    
    Runge--Kutta (RK) methods are powerful tools for numerically solving systems of first-order ordinary differential equations, and as such are often the default option in numerical routines for this task.
    There are cases however when more efficient methods are needed than Runge--Kutta, such as where the solution exhibits rapid oscillations.
    Problems classified as oscillatory are common in physics, yet the set of tools available to solve oscillatory systems efficiently is small, and problems are often treated on a case-by-case basis, using analytic approximations such as the Wentzel--Kramers--Brillouin (WKB) method \cite{BenderOrszag}.
    
    In this paper, we develop a method for a more general solution, motivated by the Mukhanov--Sasaki equation~\cite{mukhanov1992}, which governs the time-evolution of curvature perturbations in the early Universe.
    It has the form of a generalised oscillator with a time-dependent frequency and a first-order derivative term present, the frequency depending on the characteristic wavenumber of the perturbation. 
    For inference in cosmology from the Cosmic Microwave Background, it is necessary either to assume an approximate form for the primordial power spectrum of curvature perturbations, or to solve the Mukhanov--Sasaki equation for a range of characteristic wavenumbers to compute a spectrum (see, e.g.\ \cite{tasi} for a thorough review). In the event of single-field slow-roll inflation, most models lead to a scale-invariant primordial power spectrum~\cite{Dodelson_MC} which can be obtained analytically~\cite{liddle-lyth}, but models that introduce features in the primordial power spectrum can improve the fit to Cosmic Microwave Background (CMB) observations~\cite{wmap-features}. 
    In such cases when one relies on a numerical solution of the Mukhanov--Sasaki equation, in regions where the perturbation is oscillatory, this is a challenging task for Runge--Kutta-based methods.
    Runge--Kutta solvers such as \texttt{BINGO}~\cite{bingo} can prove efficient for some single-field inflation models by taking a shortcut and not integrating the perturbation throughout its oscillatory phase~\cite{Salopek89}.
    
    There has been a proposal for an algorithm in pre-print \cite{rkwkb} that generalises the Runge--Kutta stepping procedure, but uses the WKB approximation to forecast the solution instead of a Taylor expansion when the solution is highly oscillatory.
    The proposed algorithm was named RKWKB, and while it served as the theoretical foundation of our present work, there are a number of key differences. Most importantly, we extended the algorithm so that it can be applied to damped oscillators, and to equations without closed-form frequency and first-derivative terms. We also made significant adjustments to the adaptive stepsize algorithm and the method to evaluate whether the WKB approximation is applicable at the current timestep. These modifications are outlined in section \ref{sec:methods} and detailed in appendix \ref{sec:error-estimates}--\ref{sec:summary}.
    
    We present a general purpose solver for differential equations of the form 
    \begin{equation}\label{eq:diff-eq-def}
    \ddot{x}(t) + 2\gamma(t)\dot{x}(t) + \omega^2(t)x(t) = 0,
    \end{equation}
    where~$\gamma$ and~$\omega$ may or may not be expressed as a closed-form function of time. If they cannot be, but depend on time though a set of `background' variables that can be obtained numerically, they may be supplied to the solver as array-like data structures sampled over time, as detailed in Section \ref{sec:gridding}.  Since the efficiency of the solver relies on the WKB approximation being valid for a portion of the integration range, the solver is intended for problems where the frequency is slowly varying (relative to the timescales of the problem) for a part of the integration range. The numerical solver, \texttt{oscode}, is available on github \footnote{\url{https://github.com/fruzsinaagocs/oscode}}, and can be accessed via its C++ or Python interface. 
    
    This paper is structured as follows.
    In the next section we present an overview of the algorithm and the methods used therein, leaving some details to appendices.
    This is followed by applications of the method to physical systems in Section \ref{sec:results}.
    Section \ref{sec:limitations} discusses factors the user needs to be aware of that might limit the performance of the solver, as well as future improvements and extensions.
    We conclude in Section \ref{sec:conclusion} with a short summary. 
    
\section{Methods}\label{sec:methods}
    \subsection{Overview}\label{sec:overview}
        The basis for our solver is the generalised stepping approach detailed in~\cite{rkwkb}, which we will summarise here. Having a numerical estimate for the solution~$x$ and its derivative~$\dot{x}$ at time~$t$, the solution at a later time~$t+h$ is obtained. Then, using an error estimate on the proposed step, the stepsize~$h$ is updated such that the error estimate stays within a local tolerance limit. Such adaptive control of the stepsize is a requirement for robust numerical solvers. Starting from two functions~$f_{\pm}(t)$ that form an appropriate basis set for the true solution of the second-order differential equation, and are linearly independent at all~$t$, we match the correct solution and its derivative by linearly combining~$f_{\pm}$ and their derivatives:
        \begin{equation}\label{eq:gen-step-x}
         x(t+h) = A_{+}f_{+}(t+h) + A_{-}f_{-}(t+h), 
        \end{equation}
        and
        \begin{equation}\label{eq:gen-step-dx}
        \dot{x}(t+h) = B_{+}\dot{f}_{+}(t+h) + B_{-}\dot{f}_{-}(t+h),
        \end{equation}
        where
        \begin{equation}\label{eq:gen-step-a}
        A_{\pm} = \frac{\dot{x}(t)f_{\mp}(t)-x(t)\dot{f}_{\mp}(t)}{\dot{f}_{\pm}(t)f_{\mp}(t)-\dot{f}_{\mp}(t)f_{\pm}(t)},
        \end{equation}
        and 
        \begin{equation}\label{eq:gen-step-b}
        B_{\pm} = \frac{\ddot{x}(t)\dot{f}_{\mp}(t)-\dot{x}(t)\ddot{f}_{\mp}(t)}{\ddot{f}_{\pm}(t)\dot{f}_{\mp}(t)-\ddot{f}_{\mp}(t)\dot{f}_{\pm}(t)}.
        \end{equation}
        In the above,~$\ddot{x}(t)$ may be obtained from the differential equation itself, using~$x(t)$ and~$\dot{x}(t)$.
        It it shown in~\cite{rkwkb} that the above procedure reduces to Euler's method in the limit of vanishing stepsize~$h$ and with the appropriate choice of~$f_{\pm}~$.
        The above approach therefore allows one to pick trial solutions~$f_{\pm}$ that approximate the true solution well over a larger range than an~$n^{\text{th}}$~order polynomial, which would be the choice for~$f_{\pm}$ in the case of an~$n^{\text{th}}$~order Runge--Kutta method.
        
        As~\cite{rkwkb} suggests, the WKB method can be used to derive an analytic approximation to the true solution of a single oscillator on timescales much shorter than~$\frac{\omega}{\dot{\omega}}$, the timescale on which the frequency changes. The WKB solutions, detailed in the following subsection, are ideal candidates for~$f_{\pm}$ over such timescales.
        
        In general however, the frequency cannot be expected to vary slowly over the entire range of integration, and the WKB solutions might not always be a good choice for~$f_{\pm}$.
        To counter this, a dynamic switching mechanism is included in the solver, which consists of attempting two steps of size~$h$ simultaneously.
        First, a Runge--Kutta step of order 5 is calculated (a `RK step' hereafter), then a step using Equations (\ref{eq:gen-step-x})--(\ref{eq:gen-step-b}), with~$f_{\pm}$ set to the WKB solutions (a `WKB step').
        Based on the error estimates on each of these, the next stepsize,~$h^{\ast}$, is computed.
        The step with the larger next predicted stepsize is chosen.
        This is to minimise the number of steps the solver needs to take to achieve a given local accuracy, and hence minimise runtime.
        The step with the chosen method may be accepted or rejected, and the stepsize~$h$ increased or decreased. 

        The two methods are described in greater detail in the subsections that follow, with their error estimates discussed in  appendix \ref{sec:error-estimates}. Details of switching between methods and updating the stepsize can be found in appendix \ref{sec:stepping}. Finally, a step-by-step summary of the algorithm is given in appendix \ref{sec:summary}.
      
    \subsection{Wentzel--Kramers--Brillouin solutions}\label{sec:wkb}
        Starting from the equation
        \begin{equation}\label{eq:eom-wkb}
        \ddot{x}(t) + 2\gamma(t)\dot{x}(t) + \omega^2(t)x(t) = 0, 
        \end{equation}
        we wish to derive asymptotic expansions of the two independent solutions, in the limit that~$\omega$ is slowly varying relative to~$x$, but~$\gamma$ need not be so. In the absence of a first-derivative term the derivation starts by introducing a power-counting parameter~$T$, with~$T \gg 1$: 
        \begin{equation}\label{eq:eom-scaled-1}
        \ddot{x} + T^{2}\omega^2 x = 0.
        \end{equation}
        If we now insert~$\gamma$ and allow it to vary on shorter timescales than~$\omega$, the equivalent equation to consider is
        \begin{equation}\label{eq:eom-scaled-correct}
        \ddot{x} + 2\gamma \dot{x} + T^{2}\omega^2 x = 0.
        \end{equation}
        Following~\cite{BenderOrszag}, one can then seek asymptotic approximations in the form of an exponential power series \footnote{Note that the following expression appears erroneously in~\cite{rkwkb}, in that~$T$ should be replaced with~$T^{-1}$.}
        \begin{equation}\label{eq:wkb-expansion}
        x(t) \sim \exp\left( T \sum_{n=0}^{\infty} S_n(t)T^{-n} \right).
        \end{equation}
        Substituting (\ref{eq:wkb-expansion}) into (\ref{eq:eom-scaled-correct}), setting coefficients of powers of~$T$ to zero,
        one arrives at the recursion
        \begin{equation}\label{eq:s-recursion}
        \begin{split}
        \dot{S}_0(t) &= \pm i\omega, \\
        \dot{S}_i(t) &= -\frac{1}{2S_0'}\left( \ddot{S}_{i-1} + 2\gamma \dot{S}_{i-1} + \sum_{j=1}^{i-1}\dot{S}_j\dot{S}_{i-j}\right).
        \end{split}
        \end{equation}
        The first four terms in the asymptotic series in the presence of a first-derivative term are
        \begin{equation}\label{eq:s-solutions}
        \begin{split}
        S_0 &= \pm i \int\omega dt, \\
        S_1 &= -\frac{1}{2}\ln{\omega} - \int\gamma dt,\\
        S_2 &= \pm i \int -\frac{1}{2}\frac{\gamma^2}{\omega} - \frac{1}{2}\frac{\dot{\gamma}}{\omega} + \frac{3}{8}\frac{\dot{\omega}^2}{\omega^3} - \frac{1}{4}\frac{\ddot{\omega}}{\omega^2}dt, \\
        S_3 &= \frac{1}{4}\frac{\gamma^2}{\omega^2} + \frac{1}{4}\frac{\dot{\gamma}}{\omega^2} -  \frac{3}{16}\frac{\dot{\omega}^2}{\omega^4} + \frac{1}{8}\frac{\ddot{\omega}}{\omega^3}. 
        \end{split}
        \end{equation}
        As~\cite{BenderOrszag} states, the WKB series is a singular perturbative expansion. The sum in (\ref{eq:wkb-expansion}) is usually divergent (unless it truncates) and needs to be truncated at some term in order to be a good approximation to~$x(t)$.
        To use (\ref{eq:wkb-expansion}) as an approximate solution to (\ref{eq:eom-wkb}), one needs to set $T=1$. This is allowed despite having assumed $T \gg 1$ (see, e.g.\ \cite{BenderOrszag}) as long as the asymptotic inequalities
        \begin{equation}\label{eq:asymp-rels}
        TS_0(t) \gg S_1(t) \gg \ldots \gg T^{1-n}S_n(t) \\
        \end{equation}
        hold uniformly within the interval~$[t,t+h]$. If the asymptotic inequalities are satisfied and the first term not included in the asymptotic WKB series is small,
        \begin{equation}\label{eq:wkb-validity-2}
        T^{-n}S_{n+1}(t) \ll 1,    
        \end{equation}
        $x(t) \sim \exp\left( T\sum_{i=0}^n T^{i}S_i(t) \right)$ is a good approximation. 
           
        To utilise the WKB solutions, we set~$f_{\pm}(t)$ to~$x(t)$ according to (\ref{eq:wkb-expansion}) with $T=1$. 
        Computing a WKB step from~$t$ to~$t+h$ thus involves 
        \begin{equation}\label{eq:s-integral}
            S_i(t+h) - S_i(t) = \int_{t}^{t+h} \dot{S_i}(t')dt'.
        \end{equation}
        If the solver enters an integration region suitable for being approximated by WKB solutions, the stepsize~$h$ is expected to increase, and the error on the integrals (\ref{eq:s-integral}) is expected to dominate the error on~$x$ and~$\dot{x}$ in WKB steps.
        Although in these regions~$\omega$ changes slowly, care needs to be taken to evaluate the integrals accurately.
        As~$\omega(t)$ and~$\gamma(t)$ may not be available in closed form, the integrals are computed numerically. 
 
    \subsection{Numerical integration and differentiation}
        We chose to calculate the integrals (\ref{eq:s-integral}) using a method from the Gaussian quadrature family~\cite{RileyHobsonBence}, Gauss--Lobatto integration~\cite{AbramowitzStegun}.
        Gaussian quadrature formulae work by modelling the integrand as a linear combination of appropriately chosen mutually orthogonal polynomials.
        As a side effect, a certain class of integrands make the integral exact (in the Gauss--Lobatto case, polynomials of degree~$2n-3$, where~$n$ is the number of abscissas).
        Typically the remainder in such methods is proportional to a higher order ($2n-2$ for Gauss--Lobatto) derivative of the integrand, which we expect to be small in integration regions of interest, where WKB is a good approximation and several oscillations can be stepped over, such that~$h\gg \frac{2\pi}{\omega}$.
        This property makes Gaussian quadrature superior to integrating the~$\dot{S}_i$ with a Runge--Kutta step, as the latter would approximate the integral from~$t$ to~$t+h$ with a Taylor expansion around~$t$, with an error as some power of~$h$. 
        Gaussian quadrature methods are also desirable because they converge exponentially fast with~$n$, due to the order of the method increasing with~$n$ as well as the density of points of evaluation~\cite{numerical_recipes}.
        This makes them a better choice than Newton--Cotes methods with equally spaced abscissas, such as the trapezoidal rule or Simpson's method.  
        
        Gauss--Lobatto integration with~$n=6$ was chosen in particular because the abscissas it uses include the beginning and endpoints of integration,~$t$ and~$t+h$. This makes it a FSAL (first same as last) method, and one could design a~$5^{\mathrm{th}}$ order, 6-stage  Runge--Kutta formula based on the same abscissas, minimising the number of evaluations of~$\omega(t)$ and~$\gamma(t)$ during a single step of the algorithm (WKB and RK). The remainder on a Gauss--Lobatto integral is given analytically, but since it involves the~$(2n-2)^{\mathrm{th}}$ derivative of the integrand, it is more common to be estimated as the difference between the results with~$n$ and~$n-1$ abscissas.
      
        The integrands in (\ref{eq:s-solutions}) contain derivatives of~$\omega$ and~$\gamma$, which may not be available in closed form, and hence will also be calculated numerically. Since we already need to evaluate~$\omega$ and~$\gamma$ at a total of 9 distinct points (Gauss--Lobatto abscissas) for the integrals in~$S_0$ and~$S_1$, it is worth re-using these values and derive finite difference formulae using them as stencil points~\cite{jordancalculus}, by solving
        \begin{equation}\label{eq:finite-difference}
        \begin{bmatrix} w_1 \\ \vdots \\ w_n \end{bmatrix}
        =
        \frac{1}{h^D}
        \begin{bmatrix}
        s_1^0 & \ldots & s_n^0 \\
        \vdots & \ddots & \vdots \\
        s_1^{n-1} & \ldots & s_n^{n-1} \\
        \end{bmatrix}^{-1} 
        \begin{bmatrix}
        0 \\ \vdots \\ D! \\ \vdots \\ 0
        \end{bmatrix},
        \end{equation}
        where the~$s_i$ define the stencil such the points of evaluation are~$t_i = t + s_i h$,~$D$ is the order of derivative desired, and the $D!$ is the $(D+1)^{\text{th}}$ entry in the vector on the right-hand-side. The~$w_i$ are the resulting weights of the function evaluations:
        \begin{equation}\label{eq:finite-difference-ws}
        \frac{d^{D}f}{dt^D}\bigg\rvert_{t} = \sum_{i=1}^n w_i f(t_i). 
        \end{equation}
        The finite difference formulae above work by cancelling the first~$D$ terms in the Taylor expansion of~$f$ around~$t$ and setting the coefficient of the~$(D+1)^{\mathrm{th}}$ term to~$1$. These give~$D+1$ constraints, but since we are free to choose the weights of~$n_s$ evaluations of~$f$ (where~$n_s$ is the number of stencil points), we can cancel a further~$n_s-D-1$ terms in the Taylor series and thus get a result that is accurate to~$\mathcal{O}(h^{n_s-D})$. While~$h$ is expected to be large in a region of integration where WKB is a good approximation, the coefficient multiplying~$h^{n_s-D}$ is expected to be small (since the derivatives of~$\omega$ are expected to be small), and therefore we argue that finite differences is an acceptable way to estimate derivatives of~$\omega$ and~$\gamma$. 
    
    \subsection{Explicit Runge--Kutta formulae based on Gauss--Lobatto stencil points}\label{sec:rk}
        Runge--Kutta methods are a wide family of solvers that approximate the solution to a system of first-order ordinary differential equations,
        \begin{equation}
        \vec{\dot{y}(t)} = \vec{f}(\vec{y},t).
        \end{equation}
        They do so by expressing the solution at a later time~$t+h$ as
        \begin{align}
        \vec{y}(t+h) &= \vec{y}(t) + \sum_{i=1}^s b_i \vec{F}_i, \label{eq:rk-lincomb} \\ 
        \vec{F}_i &= \vec{f}(\vec{Y}_i, t + c_i h), \\ 
        \vec{Y}_i &= \vec{y}(t) + h\sum_{j=1}^s a_{ij}\vec{F}_j. \label{eq:last-rk}
        \end{align}
        For the present problem, the system to be solved is
        \begin{gather}
        \vec{y} = (x, \dot{x}) = (y_1, y_2), \\
        \vec{F} = (y_2, - \omega^2(t)y_1 - 2\gamma(t)y_2).
        \end{gather}
        
        Explicit formulae are a subset of the family for which the sum in (\ref{eq:last-rk}) on~$j$ runs until~$j < i$. These are of particular interest because the~$\vec{Y}_i$ can be calculated in an iterative manner (rather than having to solve a system of equations for them). The coefficients~$a_{ij}$,~$c_i$, and~$b_i$ fully determine the method, and can be compactly summarised in a Butcher tableau, shown in Table \ref{table:general_rk}.
        \begin{table}[h]
        \centering
        \begin{tabular}{l|lllll}
        0        &          &          &          &             &       \\
        $c_2\quad$    & $\quad a_{21}\quad $ &          &          &             &       \\
        $c_3\quad$    & $\quad a_{31}\quad $ & $\quad a_{32}\quad $ &          &             &       \\
        $\vdots\quad$ & $\quad \vdots\quad $ & $\quad \vdots\quad $ & $\quad \ddots\quad $ & &       \\
        $c_s\quad$    & $\quad a_{s1}\quad$ & $\quad a_{s2}\quad $ & $\quad \dotsm\quad $ & 
        $\quad a_{s,s-1}\quad$ &       \\ \hline
                 & $\quad b_1\quad $    & $\quad b_2\quad $    & $\quad \dotsm\quad $ & $\quad b_{s-1}\quad $   & $\quad b_s\quad $
        \end{tabular}
        \caption{Butcher tableau for an explicit Runge--Kutta method.}
        \label{table:general_rk}
        \end{table}
   
        Although there exist implicit methods with few intermediate points (or stages,~$s$) that are based on Gaussian quadrature~\cite{butcher-odes}, most Runge--Kutta formulae work on the basis of Taylor-expanding both~$\vec{y}$ and the linear combination of function evaluations on the right-hand-side of (\ref{eq:rk-lincomb}) around~$t$ by an amount~$h$, then matching coefficients of powers of~$h$ up until a given order.
        The equations resulting from counting powers of~$h$ are called order constraints, and can be derived with the help of graph theory (as detailed in~\cite{butcher-odes}).
        Particularly efficient (so-called embedded) algorithms use the same function evaluations~$\vec{F}_i$ to match coefficients to order~$N$ and~$N-1$, thus producing two estimates on~$\vec{y}(t+h)$ whose difference can be used as an error estimate.
        
        For most combinations of the number of stages~$s$ and desired order of accuracy~$N$, the order constraints do not pin down all entries in the Butcher tableau, and the leftover degrees of freedom are often fixed by minimising the coefficient of the leading-order term in the local error.
        An efficient embedded (4,5) pair developed by~\cite{bogacki-rk45} demonstrates this, and is used in {\tt rksuite}~\cite{rksuite-paper} (used by the {\tt NAG Library}~\cite{naglib}) as one of the possible Runge--Kutta formulations.
        For the present problem, we are interested in solving the order constraints of a 6-stage,~$5^{\mathrm{th}}$ order method \footnote{The highest order a 6-stage method can achieve is 5, as proven in~\cite{butcher-odes}.}, with the~$c_i$ set to the Gauss--Lobatto abscissas for~$n=6$, and a 4-stage,~$4^{\mathrm{th}}$ order method with its~$c_i$ equal to the Gauss--Lobatto abscissas for~$n=5$ with the exception of the midpoint. This way we can recycle the evaluations of~$\omega$ and~$\gamma$ at the abscissas to calculate the integrals in (\ref{eq:s-integral}), estimate their errors, take a Runge--Kutta step in~$x$,~$\dot{x}$ and get their error estimates all at the same time. 
        The order constraints for this system can be solved symbolically with no leftover degrees of freedom, demonstrated in~\cite{scicomp-maple}. The resulting coefficients are summarised in the form of Butcher tableaux in \ref{sec:rk-tableau}.
         
    \subsection{Defining~$\omega(t)$ and~$\gamma(t)$}\label{sec:gridding}
    
        In many problems of interest, the frequency and the friction term will not be explicit functions of time, but functions of variables that depend on time through a set of differential equations that may only be solved numerically.
        The algorithm requires the values of~$\omega(t)$ and~$\gamma(t)$ to be known at 9 distinct points in each step along the solution, but is otherwise blind to how the functions are defined.
        In order for the solver (and in particular Gauss--Lobatto integration) to work reliably, the frequency and friction terms need to be known at any timepoint within the integration range to high (at least 1 in~$10^{9}$) accuracy.
        
        For convenience the solver has been set up such that the user can provide values of the functions (or their natural logarithms) as vectors evaluated on an evenly spaced, monotonically increasing grid over time. It will then carry out linear interpolation whenever a function evaluation is required. The even spacing in the independent variable is a requirement for the sake of speed, as it simplifies the search for the nearest gridpoints ahead of the interpolation. 
        
        If evaluation on an evenly spaced grid is not possible or the grid cannot be made fine enough for linear interpolation to be sufficiently accurate, the user may define~$\omega(t)$ and~$\gamma(t)$ as interpolated functions using a suitable interpolation method. 
       
\section{Applications}\label{sec:results}
    \begin{figure*}
        \centering
        \includegraphics{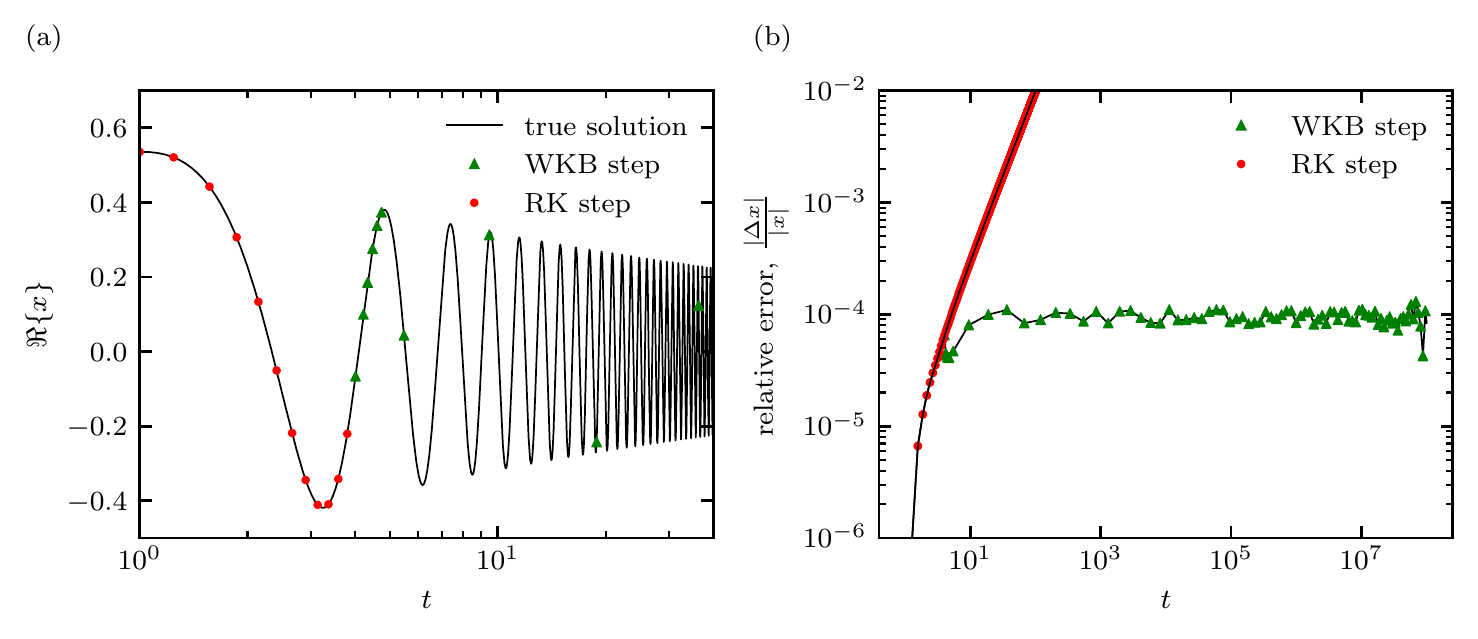}
        \caption{Numerical solution of the Airy equation obtained with the solver (dots and triangles), overlaid on the true solution as computed by the \texttt{boost} math library.
        The algorithm exhibits a clear switch from taking RK steps to WKB steps at around~$t \approx 4$, as expected.
        Despite the~$t$-axis being logarithmic, the stepsize-increase is clearly visible as time increases, and the rate of change of~$\omega$ decreases. Also shown is the accumulation of relative error during the numerical solution of the Airy equation until late times, showing the difference between a purely RK-based approach and RKWKB (\texttt{oscode}). The relative tolerance was set to be~$10^{-4}$, which the RKWKB solution does not exceed, but navigates such that the largest possible steps are taken whilst staying within this limit. In contrast, a solver taking only RK steps quickly decreases its steps whilst accumulating error.}
        \label{fig:airy-x}
    \end{figure*}
    
    \begin{figure*}
        \centering
        \includegraphics{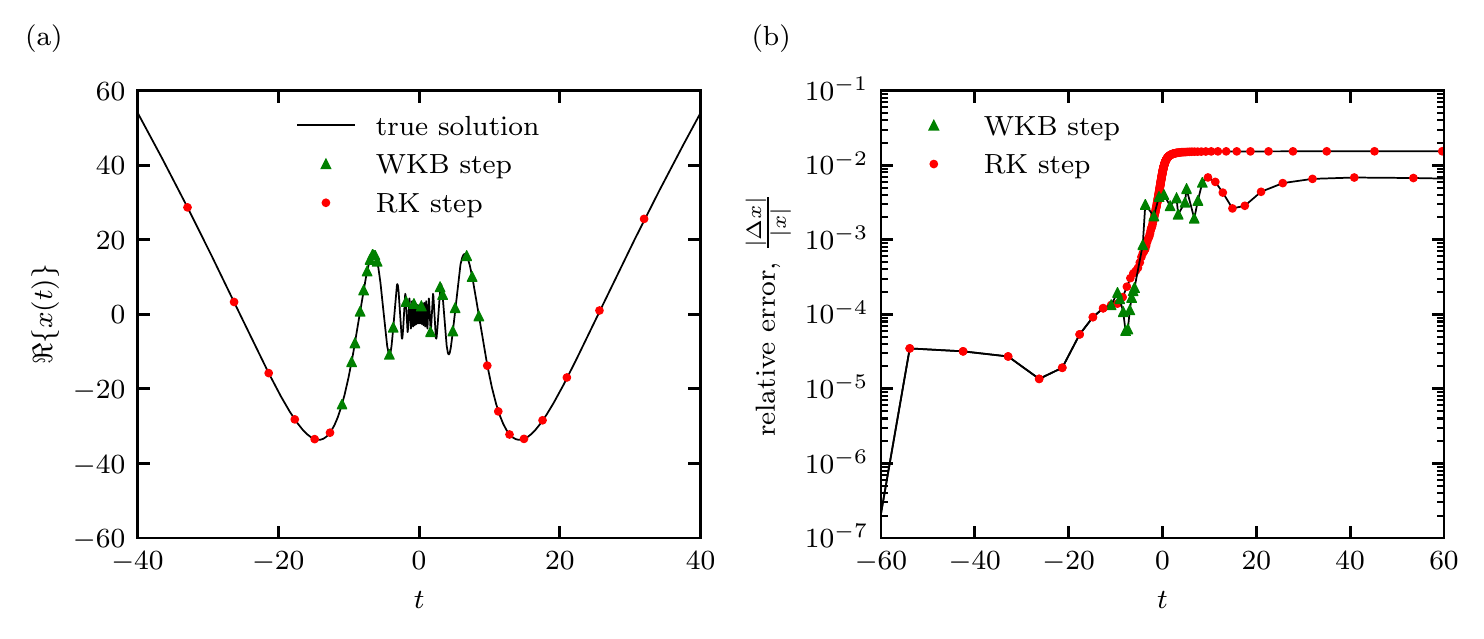}
        \caption{Numerical solution of the burst equation with~$n=40$. A relative tolerance of~$10^{-3}$ was set, and the equation was solved from~$t=-2n$ to~$t=2n$.}
        \label{fig:burst_n40}
    \end{figure*}

    \subsection{Airy equation}\label{sec:airy}
        We first demonstrate the efficiency of the solver when applied to the Airy equation,
        \begin{equation}\label{eq:airy}
        \ddot{x} + tx = 0,
        \end{equation}
        which has the solution~${\Ai(-t)+i\Bi(-t)}$. All derivatives of~$\omega$ decrease with time, hence the algorithm is expected initially to perform RK steps and at a later time switch to WKB, making the Airy equation an ideal example to test both the accuracy of the WKB steps and the stepsize-update procedure.
        This behaviour is illustrated qualitatively in Figure \ref{fig:airy-x}.
        The second panel in Figure \ref{fig:airy-x} then details the error properties of the RK and WKB phases, and shows that whilst the global relative error grows in RK steps, it levels off once the WKB phase is entered.
        In contrast, for a pure RK method, the stepsize decreases whilst the relative error continues growing. The RKWKB-based solver (\texttt{oscode}) has no difficulty stepping through the Airy solution until times as late as~$10^8$, at which point the stepsize becomes too large to store~$[S_i]_t^{t+h}$ with the required precision.
        This limitation is discussed in Section \ref{sec:limitations}.

    \subsection{Burst equation}\label{sec:burst}
    
    To illustrate the switching mechanism between RK/WKB steps, we next apply the solver to the equation
    \begin{equation}\label{eq:burst}
    \ddot{x} + \frac{n^2 -1}{(1+t^2)^2}x = 0.
    \end{equation}
    A solution for this system is 
    \begin{align}\label{burst-soln}
    x(t) = \frac{\sqrt{1+t^2}}{n}(&\cos{(n\arctan{t})} +\nonumber \\
    &+ i\sin{(n\arctan{t})}),
    \end{align} 
    characterised by a burst of approximately~$n/2$ oscillations in the region~$|t| < n$.
    The exact solution and numerical estimates of the solution at the steps taken by our solver are shown in Figure \ref{fig:burst_n40}.

        \begin{figure}[h]
            \centering
            \includegraphics{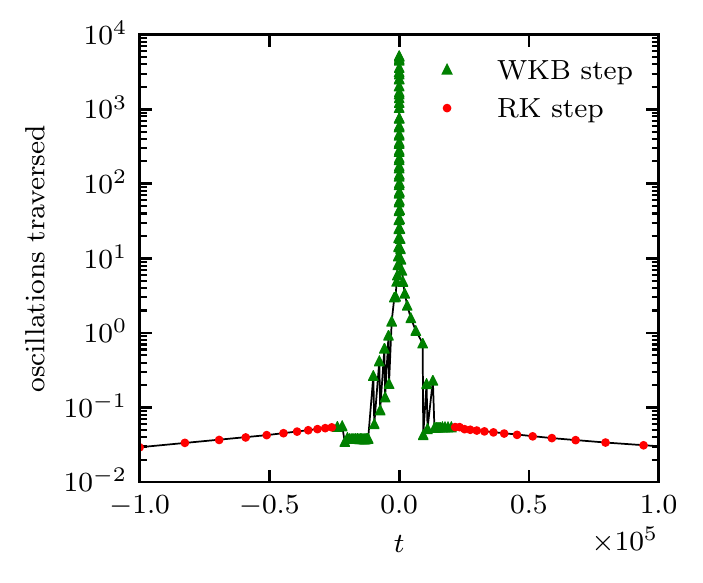}
            \caption{Number of oscillations stepped over in a single step, while solving the burst equation with~$n=10^5$, and a relative tolerance of~$10^{-4}$.}
            \label{fig:burst_oscs}
        \end{figure}      
        
        \begin{figure}[h]
            \centering
            \includegraphics{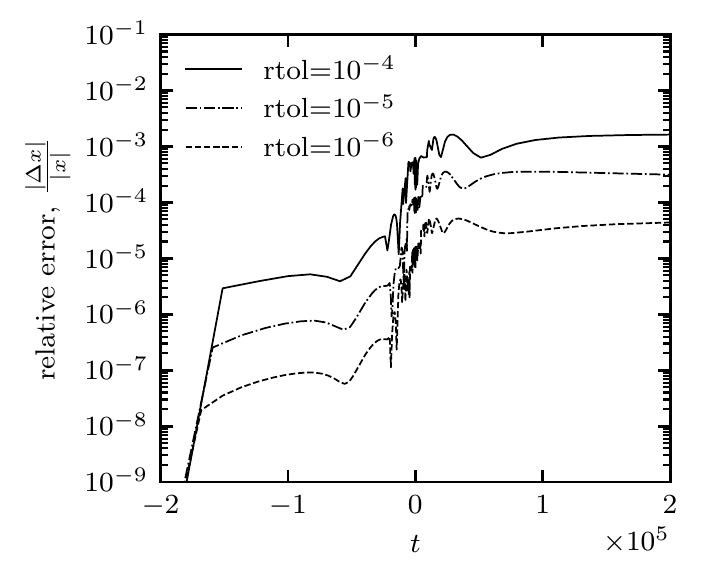}
            \caption{Progression of the relative error in the burst equation with~$n=10^5$, with different settings of the local relative tolerance `rtol'.}
            \label{fig:burst_rtols}
        \end{figure}        
       
        \begin{figure}[h]
            \centering
            \includegraphics{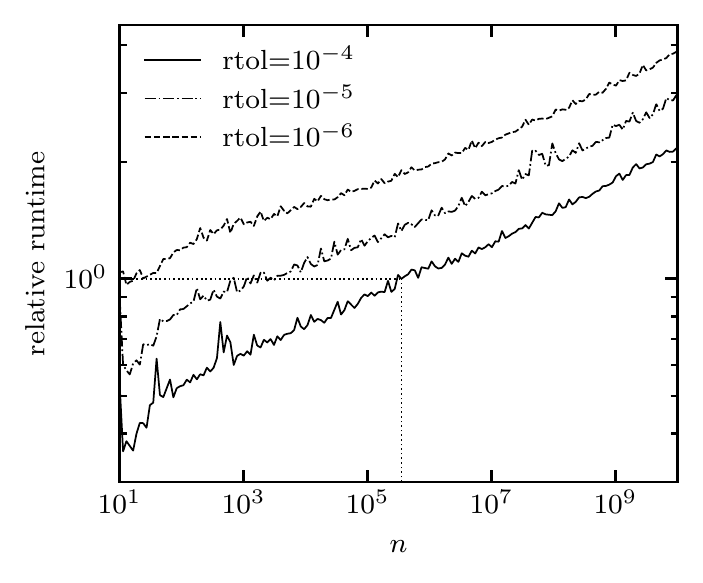}
            \caption{Relative runtime of solving the burst equation from~$t=-2n$ to~$t=2n$, with~$n$ varying from~$10^1$ to~$10^{10}$, and the relative tolerance, `rtol' from~$10^{-4}$ to~$10^{-6}$. The runtimes are referenced to the median of the~$n$-range and relative tolerance of~$10^{-4}$, as indicated by the dotted lines.}
            \label{fig:burst_runtime}
        \end{figure}       
        
        \begin{figure}[h]
            \centering
            \includegraphics{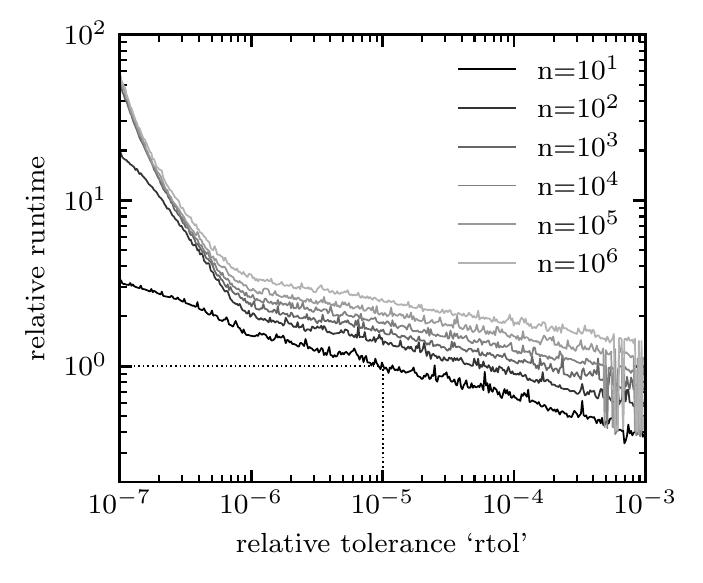}
            \caption{Relative runtime of solving the burst equation, as a function of the relative tolerance set (with the absolute tolerance set to 0). The different curves show different values of~$n$, proportional to the total number of oscillations traversed.}
            \label{fig:burst_runtime_rtol}
        \end{figure}       
        
        \begin{figure}[h]
            \centering
            \includegraphics{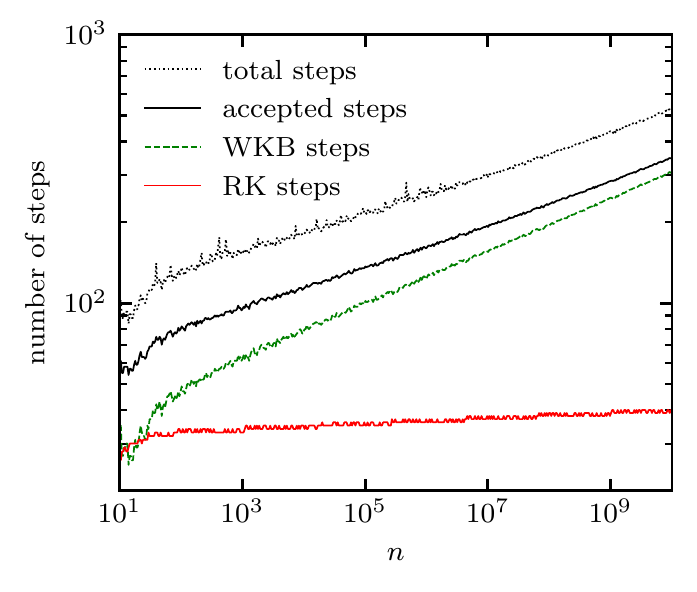}
            \caption{Step breakdown in solving the burst equation from~$t=-2n$ to~$t=2n$, with~$n$ varying from~$10^1$ to~$10^{10}$, and the relative tolerance, `rtol' set to~$10^{-4}$.}
            \label{fig:burst_steps}
        \end{figure}  
        
        Figure \ref{fig:burst_n40} also shows the error accumulated in the numerical solution of this example.
        This clearly shows that once the burst of oscillations is encountered, taking WKB steps becomes more efficient, and the solver allows the stepsize to grow until the local error reaches its tolerance limit. 
        It then keeps the local error at this limit whilst traversing as many oscillations as possible. The global error is also seen to level off, at a slightly higher value than the local tolerance.
        To demonstrate that as many oscillations are stepped over as possible, Figure \ref{fig:burst_oscs} shows the number of oscillations traversed during a single step of the solver as a function of time having a sharp peak near~$t=0$, where it is able to leap through~$10^4$ oscillations.
  
        The robustness of the algorithm was tested by monitoring the numerical error as a function of time for relative tolerances ranging from~$10^{-6}$ to~$10^{-4}$, shown in Figure \ref{fig:burst_rtols}. The global error in all of the above examples reaches a constant value of~$\sim 10\times \mathrm{rtol}$ by the end of the oscillatory phase.
        
        Finally, we show that the algorithm is efficient over a range of values of~$n$ (which determine the total number of oscillations) and tolerances in Figures \ref{fig:burst_runtime} and \ref{fig:burst_runtime_rtol}. The algorithm shows a slow, 4-fold runtime increase over 9 orders of magnitude change in the number of oscillations, which is due to the increase in WKB steps needed to traverse the oscillatory region, shown in Figure \ref{fig:burst_steps}. Figure \ref{fig:burst_runtime_rtol} also reveals that the algorithm is most efficient in the relative tolerance range of~$10^{-6}$ --~$10^{-4}$. For tolerances lower than this, a 4-5${}^{\mathrm{th}}$ order RK pair is not generally recommended. 

    \subsection{Schr\"odinger equation}\label{sec:schrodinger}

        \begin{figure*}
            \centering
            \includegraphics{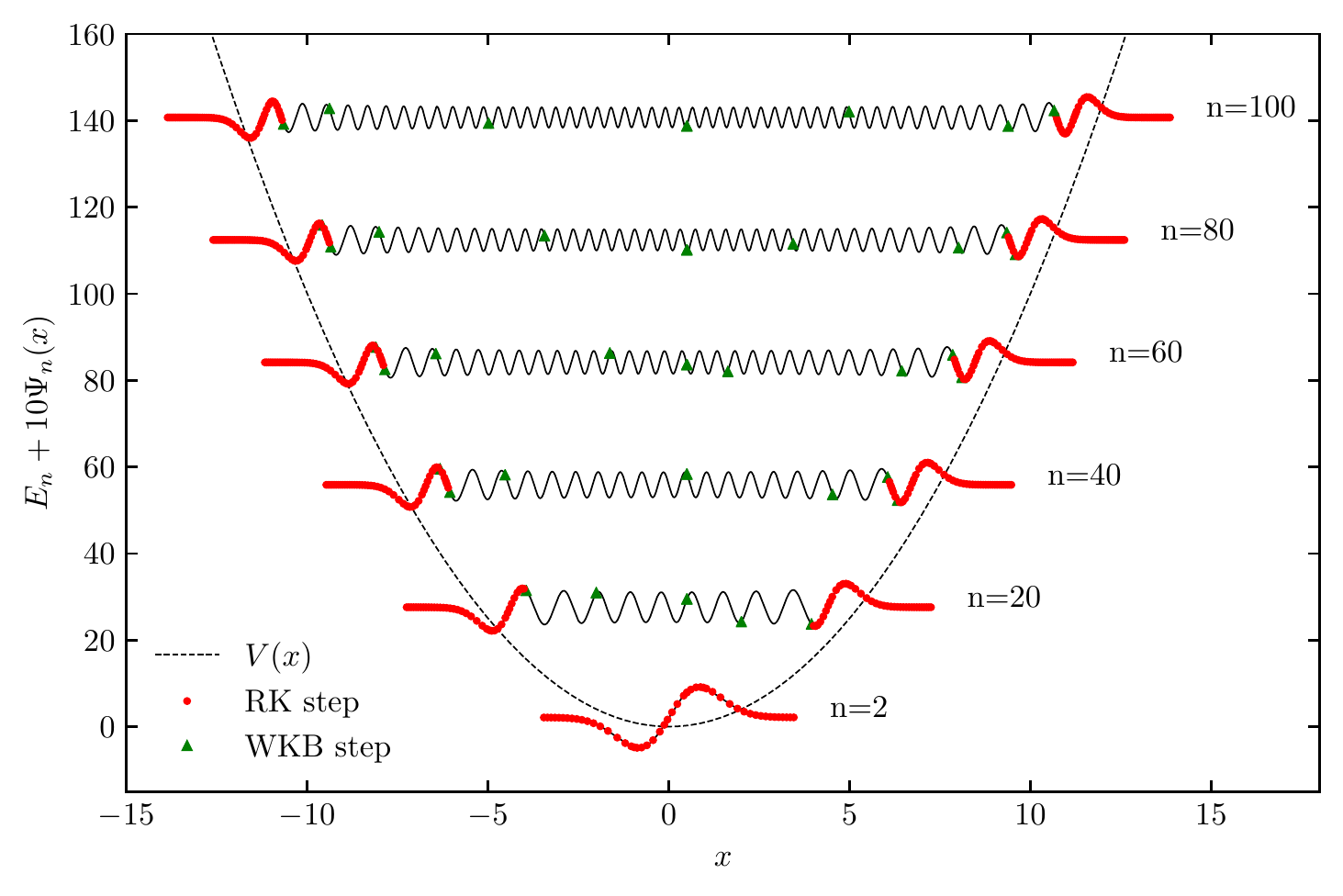}
            \caption{Energy eigenfunctions in a harmonic potential well. In units of~$m=\hbar=1$ and with a potential~$V(x)=x^2$, the~$n^{\mathrm{th}}$ level has energy~$\sqrt{2}(n - 1/2)$. The wavefunctions in this potential are given analytically in terms of the Hermite polynomials, and are plotted in black. Numerical integration was started from both sides of~$x=0$, from well outside the potential (where~$E \ll V(x)$), until~$x=0.5$. The initial conditions were set using the analytic solution for~$\Psi$ and~$\Psi'$. The relative tolerance was set to be~$10^{-3}$. }
            \label{fig:schrodinger-harmonic}
        \end{figure*}   
    
        The one-dimensional time-independent Schr\"odinger equation for a potential~$V(x)$ takes the form
        \begin{equation}\label{eq:schrodinger-1d}
        \Psi''(x)  + 2m(E-V(x))\Psi(x) = 0,
        \end{equation}
        where we set~$\hbar=1$.
        The WKB method's original use was to compute approximate solutions of (\ref{eq:schrodinger-1d}), which suggests that our solver can be used as an alternative to traditional methods (such as the Numerov method~\cite{numerov}) to calculate fast numerical solutions.
        Starting with an analytic example, Figure \ref{fig:schrodinger-harmonic} shows the numerical evaluation of the energy eigenfunction~$\Psi_n$ for the~$n^{\mathrm{th}}$ energy level in a harmonic potential well, for a range of~$n$-s including high-energy excited states. Figure \ref{fig:schrodinger-harmonic} clearly shows that \texttt{oscode} only needs to take a few steps once inside the potential well, suggesting that computation time is greatly reduced relative to purely Runge--Kutta based approaches. In this example the analytic solution for the eigenfunctions were available and were used to set the values of~$\Psi$ and~$\Psi'$ at the integration boundaries. 
        
        In a general potential well, analytic solutions are not accessible and the energy eigenvalues are unknowns to be computed. Shooting methods~\cite{killingbeck-shooting} are frequently used to estimate the eigenvalues in such cases. We employ one such method to find the energies of the quantum harmonic oscillator with quartic anharmonicity, which has the potential
        \begin{equation}\label{eq:quant-anharmonic}
        V(x) = x^2 + \lambda x^4. 
        \end{equation}
        An initial guess for the eigenvalue,~$E$, is made. We start integration from points~$\pm x_0$ outside the potential on either side of~$x=0$, where~$E \ll V(x)$, using the initial conditions~$\Psi(\pm x_0)=0$ and~$\Psi'(\pm x_0)=1$. We integrate towards the inside of the potential well in order to avoid contamination of the exponentially decaying solution by the growing mode when one integrates away from the well. The first initial condition is a good approximation far outside the potential well, and~$\Psi'$ can be chosen arbitrarily as it accounts to a choice of normalisation. The two numerical solutions,~$\Psi_L$ and~$\Psi_R$ meet at an intermediate point~$x_1$. At~$x_1$, both~$\Psi$ and~$\Psi'$ must be continuous if~$E$ is an eigenvalue. Therefore the normalisation-independent quantity
        \begin{equation}
        \frac{\Psi'_L}{\Psi_L} - \frac{\Psi'_R}{\Psi_R} 
        \end{equation}
        is minimised as a function of~$E$. A few examples of the eigenvalues thus computed are presented in Table \ref{table:anharmonic-spectrum}, alongside their matching values from~\cite{banerjee}. Note that in order to get equivalent eigenvalues, we set~$m=0.5$. The eigenvalues are in good agreement up to highly excited states.
        
        \begin{table}[h]
        \centering
        \footnotesize{
        \begin{tabular}{l|cc}
        $n$ & $E_n$ & $E_n^{\ast}$\\ \hline
        $0$ & $ 1.392353 $ & 1.392352 \\
        $1$ & $ 4.648815 $ & 4.648813 \\
        $2$ & $ 8.6550501 $ & 8.6550500\\
        $3$ & $ 13.156806 $ & 13.156804 \\
        $4$ & $ 18.0577 $ & 18.0576 \\
        $15$ & $ 88.6104 $ & 88.6103\\
        $16$ & $ 96.1291 $ & 96.1296 \\
        $17$ & $ 103.793 $  &103.795 \\
        $18$ & $ 111.6025 $ & 111.6020\\
        $19$ & $ 119.5440 $ & 119.5442\\
        $50$ & $ 417.05620 $ & 417.05626\\
        $100$ & $ 1035.5440 $ & 1035.5442\\
        $1000$ & $ 21932.7848 $ & 21932.7840\\
        $10000$ & $ 471103.81 $ & 471103.80\\
        \end{tabular}
        }
        \caption{Energy eigenvalues of the quantum harmonic oscillator with quartic anharmonicity. The left-hand column~$E_n$ shows the eigenvalues found with our method, to be compared with the right-hand column~$E_n^{\ast}$, which lists the (rounded) results of~\cite{banerjee}.}
        \label{table:anharmonic-spectrum}
    \end{table}    
       
    \subsection{Mukhanov--Sasaki equation}\label{sec:ms}
        
        In the previous two toy examples, there was only a frequency,~$\omega$-term present in the differential equation to be solved, and it was available to arbitrary precision.
        This may not always be the case, as (1) one may want to switch to a more physically meaningful independent-dependent variable pair, which can introduce a friction term~$\gamma$, and (2) the frequency and friction terms might themselves be available only through numerically solving a set of differential equations. The Mukhanov--Sasaki equation illustrates both of these cases. In the brief introduction to the background of the equation to follow, we use Planck units
        \begin{equation*}
        c = \hbar = k_{\mathrm{B}} = G = 1,
        \end{equation*}
        and set the Planck mass to one,~$m_P = 1$.
        
        The Mukhanov--Sasaki equation describes the time-evolution of perturbations in a homogeneous, isotropic, `background' universe. This background, in the simplest models, assumes the presence of a single time-dependent scalar field~$\phi(t)$ (the inflaton field). The field has self-interactions described by the potential~$V(\phi)$, and its dynamics are defined by the action
        \begin{equation}\label{eq:mukhanov-action}
        \mathcal{S} = \int d^4x\sqrt{-g}\Big( \frac{1}{2}R + \frac{1}{2}g^{\mu\nu}\partial_{\mu}\phi\partial_{\nu}\phi - V(\phi) \Big).
        \end{equation}
        Assuming a metric of the Friedmann--Robertson--Walker form, the above action leads to the equations of motion
        \begin{equation}\label{eq:friedmann-1}
        H^2 + \frac{K}{a^2} = \frac{1}{3}\Big(\frac{1}{2}\dot{\phi}^2 + V(\phi)\Big),
        \end{equation}
        \begin{equation}\label{eq:friedmann-2}
        \dot{H} + H^2 = -\frac{1}{3}\Big(\dot{\phi}^2 - V(\phi) \Big),
        \end{equation}
        \begin{equation}\label{eq:bg-cty}
        0 = \ddot{\phi} + 3H\dot{\phi} + V_{,\phi},
        \end{equation}
        out of which only two are independent.
        In the above,~$a(t)$ is the scale factor,~$H(t)$ is the Hubble parameter defined as~$H = \frac{\dot{a}}{a}$, and~$K$ is the curvature, taking values~$0$,~$\pm1$ for flat, closed and open universes.
        In this section we consider flat and closed universe models, starting with the flat case. In what follows,~$K=0$ until stated otherwise.
        Perturbing the field and the metric, and introducing the gauge-invariant scalar~$\mathcal{R}$ (called the comoving curvature perturbation) one can then arrive at the Mukhanov--Sasaki equation, which we write as
        \begin{equation}\label{eq:ms-R}
        \ddot{\mathcal{R}}_k + 2\left( \frac{\ddot{\phi}}{\dot{\phi}} - \frac{1}{2}\dot{\phi}^2 + \frac{3}{2} \right)\dot{\mathcal{R}}_k + \left( \frac{k}{aH} \right)^2\mathcal{R}_k = 0.
        \end{equation}
        In the above equation, the overdot denotes differentiation with respect to~$N = \ln{a}$, and
~$\mathcal{R}_k$ is the mode with wavenumber~$k$ in the Fourier decomposition of~$\mathcal{R}$.~$N$ measures the amount of expansion the universe goes through. Even in this simple model, the single scalar field~$\phi$ is enough to trigger an accelerated expansion of the universe, inflation (\cite{Guth1982},~\cite{liddle-lyth}). During inflation,~$a(t) \sim e^{Ht}$ and~$H$ is approximately constant, hence~$N$ is a natural independent variable candidate. Another important characteristic of inflation is that the quantity~$\frac{1}{aH}$, called the comoving Hubble horizon, shrinks. The Hubble horizon plays a crucial role in governing the dynamics of perturbations, which will be described later. 
        
        In the limit of \emph{slow-roll} inflation, defined by~$\frac{1}{2}\dot{\phi}^2 \ll V(\phi)$, the background equations (\ref{eq:friedmann-1})--(\ref{eq:bg-cty}) admit analytic solutions.
        They also do in the opposite limit,~$\frac{1}{2}\dot{\phi}^2 \gg V(\phi)$, called \emph{kinetic dominance} (see~\cite{contaldi-kd}).
        Kinetic dominance has been shown to be the limit the universe emerges from in most single-field models~\cite{kineticic}. In kinetic dominance the comoving Hubble horizon grows, then shrinks again as slow-roll inflation is entered.
        Both limits can thus be used to set initial conditions to equations (\ref{eq:friedmann-1})--(\ref{eq:bg-cty}), which can then be integrated numerically. 
        
        The Mukhanov--Sasaki equation in the flat case can also be solved analytically if for all~$k$-modes of interest,~$k \gg aH$.
        Since~$k^{-1}$ is the characteristic lengthscale of a perturbation mode, this means that all modes of interest are assumed to be well inside the Hubble horizon.
        Letting the Mukhanov--Sasaki equation emerge from this limit is equivalent to choosing a vacuum state (see~\cite{birrell-davies}) which, together with a normalisation condition, are enough to provide initial conditions for the mode functions~$\mathcal{R}_k$.
        This choice of vacuum and the initial conditions are referred to as \emph{Bunch--Davies}.
        With a different choice of vacuum, it is possible to set initial conditions on the~$\mathcal{R}_k$ in kinetic dominance, when modes are not necessarily inside the Hubble horizon.
        The form of~$\mathcal{R}_k$ in the kinetically dominated limit are derived in~\cite{nqicfi}.
        In the models investigated, we shall consider both slow-roll and kinetically dominated initial conditions for the background and the perturbations.

        The Mukhanov--Sasaki equation (\ref{eq:ms-R}) is of the form of a generalised oscillator with a first-derivative~$\gamma$ term present, with both~$\gamma$ and the frequency~$\omega$ being (in general non-analytic) functions of time as they depend on the cosmological background. It follows that when a~$k$-mode is inside the Hubble horizon,~$k > aH$, it oscillates with some varying amplitude and frequency usually proportional to~$k$, and one can show that the mode `freezes out' once outside the Hubble horizon, meaning~$\mathcal{R}_k \sim \mathrm{const}$. The wavenumber-dependence of the frequency term makes this equation challenging to solve for large values of~$k$ without resorting to approximations.       
        
        Our goal is to solve the Mukhanov--Sasaki for a range of~$k$-modes until each mode has a constant amplitude, up to large values of~$k$, in order to obtain the primordial power spectrum
        \begin{equation}\label{eq:pps-rk}
        \mathcal{P}_{\mathcal{R}}^{2}(k)=\frac{k^{3}}{2 \pi^{2}}\left|\mathcal{R}_k\right|^{2}.
        \end{equation}
        
        \subsubsection{Comparison with \texttt{BINGO}}\label{sec:bingo} 
        
        We shall first adopt the computational strategy employed by many solvers designed to compute primordial power spectra, for example \texttt{BINGO}~\cite{bingo}, and \texttt{ModeCode}~\cite{PeirisModeCode}, and compare our solver performance with the former. \texttt{BINGO} is a Fortran-based code for efficient evaluation of the scalar bi-spectrum, that has to calculate the primordial power spectrum of scalar perturbations on the way, but we shall only use it to compute the primordial power spectrum.

\begin{figure}[h]
            \centering
            \includegraphics{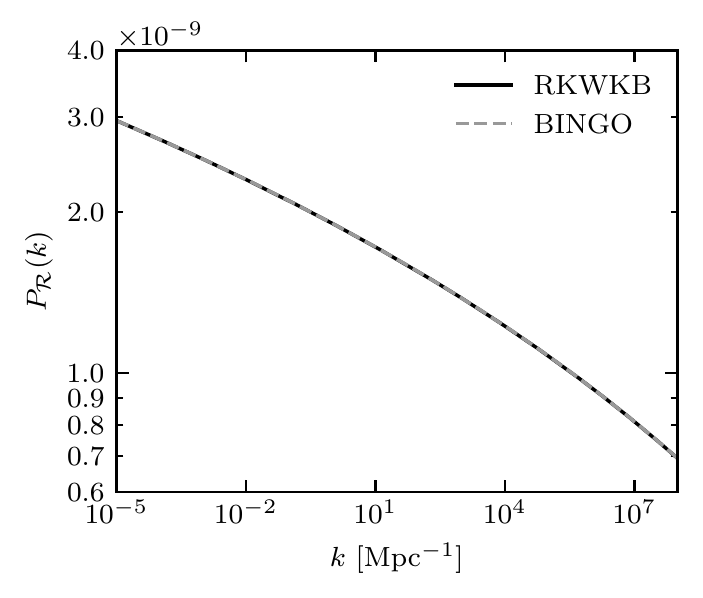}
            \caption{Primordial power spectra of the gauge-invariant scalar curvature perturbations, generated by \texttt{BINGO} and our solver with identical parameters. The spectra have been computed up to extremely large values of~$k$ for the sake of comparing the runtimes of the two codes.}
            \label{fig:rkwkb-bingo-pps}
        \end{figure}

        \texttt{BINGO} gets around the computational challenges by using a trick: it has been shown that in the case of a single-field inflationary model and assuming the universe emerges from slow-roll inflation, it is sufficient to evolve each curvature perturbation from a time they are well inside the Hubble horizon  (from, say,~$k/aH = 100$, see~\cite{Salopek89}), until the perturbation freezes out outside of the Hubble horizon ($k/aH = 10^{-2}$). This avoids integrating the solution through the majority of its oscillatory phase.
        For the comparison to be fair, we will do the same.
        First, the cosmological background ($\phi(N)$,~$\dot{\phi}(N)$,~$\ldots$) is computed numerically as a function of~$N$, starting from the slow-roll conditions, set such that the total number of e-folds of inflation,~$N_{\mathrm{tot}}\sim 60$.
        The inflationary model used in this example involves a quadratic potential, 
        \begin{equation}\label{eq:infl-pot}
        V(\phi) = \frac{1}{2}m^2\phi^2,
        \end{equation}
        where we set the inflaton mass to one,~$m=1$.
        The initial scale factor is set such that a pivot mode, corresponding to~$k=0.05$ Mpc${}^{-1}$ leaves the Hubble horizon when there are 50 e-folds of inflation left.
        For each mode, we find the~$N$ corresponding to the start and end of integration, then for our solver, we supply the algorithm with~$\omega(N)$ and~$\gamma(N)$ defined as grids, on which we perform linear interpolation (the grid needs to be sufficiently fine - for the present example we used~$5\times 10^5$ equally spaced points between~$N=0$ and~$N=75$).
        We then solve the mode evolution for each~$k$ starting from Bunch--Davies initial conditions.
        We set the same parameters to \texttt{BINGO} and our solver, in particular we set a relative tolerance of~$10^{-4}$ and an absolute tolerance of~$0$. The resulting power spectra are identical, as shown in Figure \ref{fig:rkwkb-bingo-pps}.

\begin{figure}[h]
             \centering
             \includegraphics{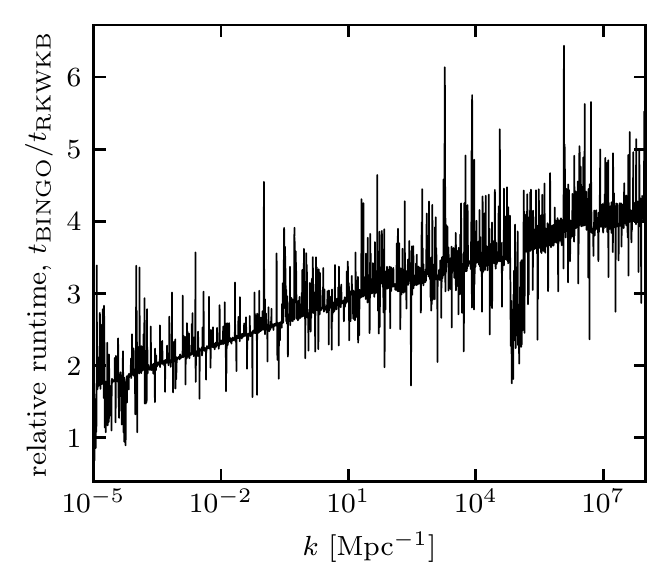}
             \caption{Ratio of the runtime of \texttt{BINGO} and our solver during the evaluation of a scalar primordial power spectrum, as a function of wavevector. }
             \label{fig:rkwkb-v-bingo-reltimes}
         \end{figure}

\begin{figure}[h]
             \centering
             \includegraphics{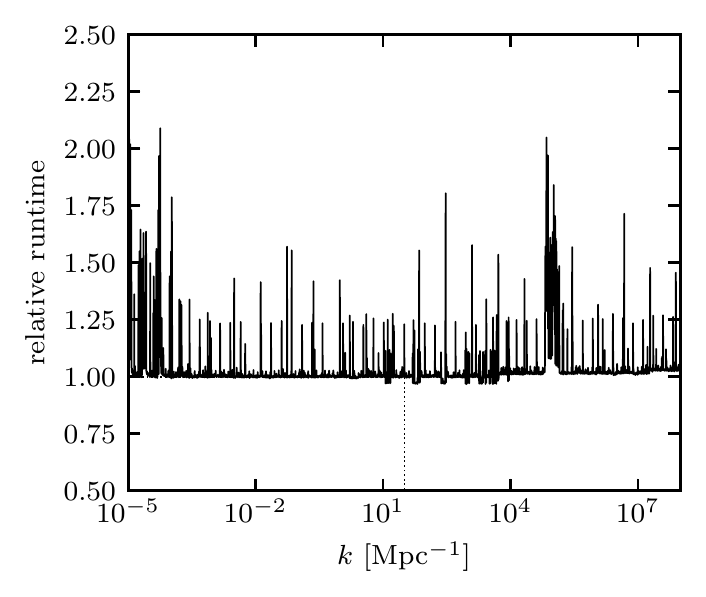}
             \caption{Progression of our solver's relative runtime with increasing wavenumber, whilst calculating a scalar primordial power spectrum. The times are referenced to the computation time corresponding to the median~$k$-value, indicated by the dotted lines. }
             \label{fig:rkwkb-reltimes}
         \end{figure}

        The computation time for the solver to obtain~$\mathcal{R}_k$ is measured and plotted as a function of~$k$ in Figures \ref{fig:rkwkb-v-bingo-reltimes} and  \ref{fig:rkwkb-reltimes}. The former shows the ratio of \texttt{BINGO} and our solver's runtimes as a function of~$k$, and the latter just that of our solver, relative to the median~$k$. Together they show that \texttt{BINGO}'s runtime is logarithmic in~$k$, whereas our solver's is constant. They also show that \texttt{oscode} performs better than \texttt{BINGO} by at least a factor of two, and at most a factor of 4 in the~$k$-range of interest.
         
\begin{figure}[h]
            \centering
            \includegraphics{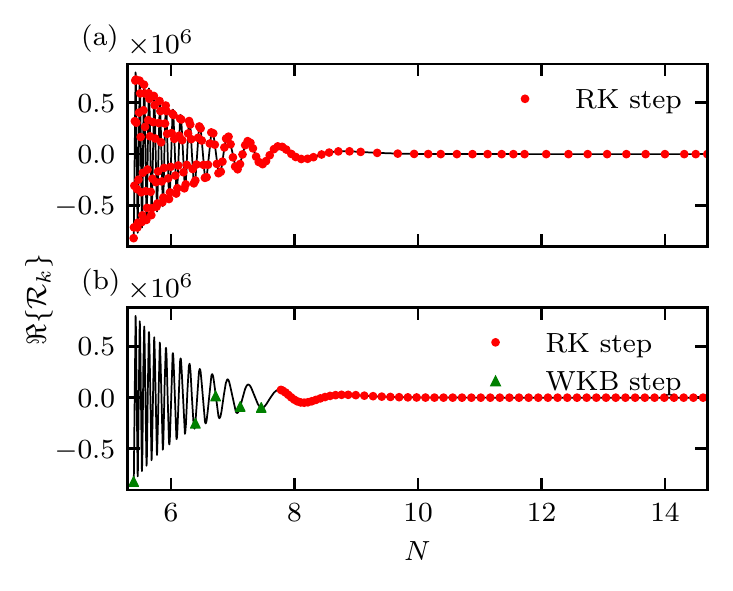}
            \caption{Comparison of \texttt{BINGO} and our solver in the evolution of a  single perturbation with wavevector~$k=10^{-5}$ Mpc${}^{-1}$. The black reference line is a dense solution generated with a Runge--Kutta (7,8)${}^{\mathrm{th}}$ order pair.
            On top of it the top panel shows the steps that \texttt{RKSUITE}'s (4,5)${}^{\mathrm{th}}$ order Runge--Kutta solver takes (a total of~$\sim 150$), the bottom panel the steps that our solver takes (a total of~$\sim 60$).
            The relative tolerance was set to~$10^{-4}$ for both methods.}
            \label{fig:ms_single_k}
        \end{figure}

        This can be explained by looking at (\ref{eq:ms-R}).
        The frequency term is a fixed number at the start and end of integration, so it will no longer scale with~$k$.
        The friction term during inflation is approximately constant.
        The range of integration,~$\Delta N$, is determined by the points where~$k/aH = c_0~$ (a constant), which during inflation is also roughly constant. 
        Therefore the number of oscillations over the range of wavenumbers in the spectrum barely changes, and we expect a WKB-based method to traverse the oscillations in constant time. In reality, the integration range increases slowly with~$N$, and small variations in the friction term cause the oscillations to change in shape, hence the slow increase in the runtime of \texttt{BINGO}. The two-fold runtime-difference present even at the smallest values of~$k$ can be explained by the difference in the number of steps taken.
        Figure \ref{fig:ms_single_k} shows the intermediate steps taken by \texttt{RKSUITE}, a numerical routine implementing efficient Runge--Kutta methods and used by \texttt{BINGO}, and the intermediate steps taken by \texttt{oscode}, whilst computing the time-evolution of a single~$k$-mode. \texttt{oscode} is able to traverse the oscillatory region of the mode's evolution in significantly fewer steps than the Runge--Kutta method, giving a reduction in computing time.

        In models where one has to start integrating the mode equation from deeper within the horizon, the starting frequency during the evolution of modes is larger, and the performance difference between an RKWKB-based approach and a Runge--Kutta integrator is even more distinct.
        Examples include universes emerging from kinetic dominance, axion monodromy models~\cite{axion-monodromy} or models with alpha vacua initial conditions~\cite{alpha-vacua}.

        \subsubsection{A model using kinetic dominance}\label{sec:kd}
       
        Inflationary models including kinetic dominance are already being investigated, e.g.\ by~\cite{hergt-kd-short} and~\cite{hergt-kd-long}.
        In this scenario, the cosmological background in terms of~$N$ is integrated from the initial state 
        \begin{align}
            \phi &= \phi_P - \sqrt{6}\ln N, \\
            \dot{\phi} &= -\sqrt{6 - \frac{2V}{H^2}}, \\
            H &= \frac{1}{3}e^{-3N},
        \end{align}\label{eq:kinetic-bg}
        where~$\dot{\phi}$ contains a contribution from the potential in order to make the system numerically stable. In kinetic dominance,~\cite{nqicfi} obtains a solution for the perturbation modes, which in terms of~$N$ take the form
        \begin{align}
        \mathcal{R}_k = \sqrt{\frac{3\pi}{8}}\frac{1}{z}e^N \Big[ &A_k H^{(1)}_0\left(\frac{3}{2}ke^{2N}\right) + \nonumber \\
        + & B_k H^{(2)}_0\left(\frac{3}{2}ke^{2N}\right) \Big], \\
        \dot{\mathcal{R}}_k = \sqrt{\frac{27\pi}{8}}\frac{k}{z}e^{3N}\Big[ &A_k H^{(1)}_0{}'\left(\frac{3}{2}ke^{2N}\right)  + \nonumber\\
        + & B_k H^{(2)}_0{}' \left(\frac{3}{2}ke^{2N}\right)\Big] + \nonumber\\
        + &\left( -\frac{\dot{z}}{z} + 1 \right)\mathcal{R}_k ,
        \end{align}
        where~$H^{(1)}$ and~$H^{(2)}$ are Hankel functions of the first and second kind, and~$A_k$,~$B_k$ are constants. We set~$A_k=0$ and~$B_k=1$, and chose parameters such that the total number of e-folds during inflation,~$N_{\text{tot}} \approx 60$, and the pivot scale corresponding to~$k=0.05$ Mpc${}^{-1}$ today leaves the horizon when there are~$N_{\ast}\approx 54$ e-folds of inflation left. We set the initial conditions for the background at~$N=0$, and for the modes at a constant~$N=1.1$, and integrate until far after horizon exit, as in Section \ref{sec:bingo}.
 
\begin{figure}[h]
            \centering
            \includegraphics{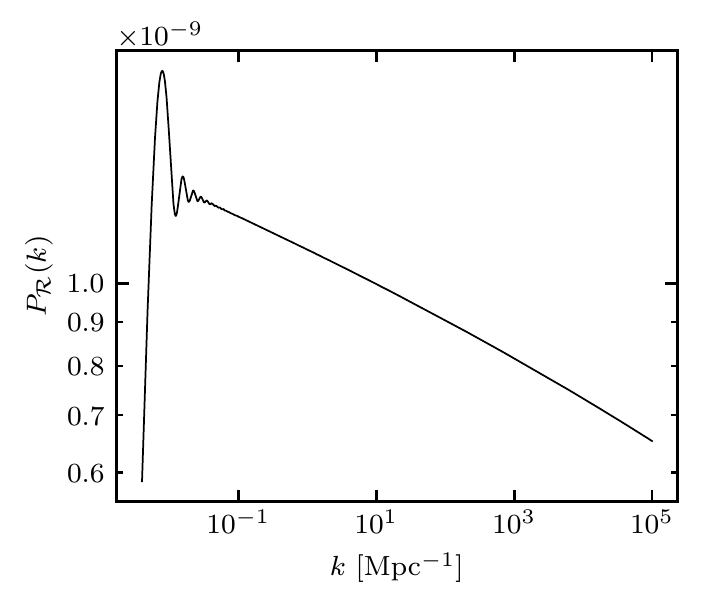}
            \caption{Scalar primordial power spectrum of perturbations emerging from kinetic dominance. The mode equation was solved from a fixed, early time (well inside kinetic dominance) until long after horizon crossing, which is only feasible if the solver used is capable of traversing many oscillations at once. The relative tolerance was set to be~$10^{-4}$. }
            \label{fig:kd-pps}
        \end{figure}

        The resulting primordial power spectrum is shown in Figure \ref{fig:kd-pps}. Such computations are only possible if the solver used can trace oscillations in the solution extremely efficiently, and indeed we found that calculating a spectrum starting from kinetic dominance and from a fixed fraction of the horizon in slow-roll can be carried out on similar timescales using our solver. 
        It is worth noting that a fast solver has been developed specifically for the Mukhanov--Sasaki equation~\cite{Haddadin} that works on the basis of using analytic approximations for when the frequency is well-approximated by an exponential or first-order polynomial.
        This gives a significant speed-up over Runge--Kutta methods, but relies on the Mukhanov--Sasaki equation to be transformable to a form without a first-order derivative term. Closed universe models do not have this property, but can still be investigated with our method.
 
        \subsubsection{A closed universe model}
        
        In this example we investigate closed universe models with curvature~$K=1$. The cosmological background evolution equations (\ref{eq:friedmann-1})--(\ref{eq:bg-cty}) can be cast into a system of linear ODE-s,
        \begin{equation}\label{eq:closed-bg-1}
        \frac{d\ln{|\Omega_k|}}{dN} = 4 + |\Omega_k|\big(4K - 2a^2V(\phi)\big),
        \end{equation}
        \begin{equation}\label{eq:closed-bg-2}
        \Big(\frac{d\phi}{dN}\Big)^2 = 6 + |\Omega_k|\big(6K - 2a^2V(\phi)\big),
        \end{equation}
        where~$\Omega_k = \frac{K}{(aH)^2}$.
        We shall consider a cosmological background emerging from kinetic dominance, such that the Hubble horizon,~$(aH)^{-1} = \sqrt{\Omega_k}$, grows until it reaches a maximum~$\sqrt{\Omega_k^i}$ at e-folds~$N_i$.
        From this point the horizon shrinks, and inflation starts.
        The parameters~$(\Omega_k^i, N_i)$, together with the requirement~$\dot{\Omega}_k(N_i) = 0$ fully fix the background evolution, and hence determine the amount of inflation,~$N_{\mathrm{tot}}$.
        We used Brent's method of root finding~\cite{brent-method} to search for the~$N_i$ for a given~$\Omega_k^i$ that yields~$N_{\mathrm{tot}}=60$.
        Hence the primordial power spectra have all other parameters fixed, with only~$\Omega_k^i$, the initial curvature at the start of inflation, changing. Integration of the background is started from~$N_i$ and is performed forwards until the end of inflation (and backwards, if necessary) to cover the integration range of the perturbation modes.
       
        The mode functions obey the generalised Mukhanov--Sasaki equation in the presence of non-zero curvature~$K$, with frequency and first-derivative terms given by~\cite{MS-general}
        \begin{align}\label{eq:ms-curved}
        \omega^2 &= \Omega_k \Big( (k_2 - K) - \frac{2Kk_2}{EK + k_2}\frac{\dot{E}}{E}\Big), \\
        2\gamma &= K\Omega_k + 3 - E + \frac{k_2}{EK + k_2}\frac{\dot{E}}{E},
        \end{align}        
        where~$ E = \frac{1}{2}\dot{\phi}^2$ and 
        \begin{equation} 
        k_2 = \begin{cases}
        k(k+2) - 3K, \; \mathrm{if} \, K > 0, \\
        k^2 - 3K, \; \mathrm{otherwise.} 
        \end{cases}
        \end{equation} 
        The modes are started from~$N = N_i$ using the Bunch--Davies conditions introduced at the start of Section \ref{sec:results}. Although the Bunch--Davies solution has been derived from the Mukhanov--Sasaki equation in a flat universe, its closed universe equivalent is not yet known. An important feature of closed universe primordial power spectra is that the values of the \emph{comoving wavenumber}~$k$, appearing in the above equations, are quantised to only take integer values, with the lowest possible value of~$k=3$~\cite{lasenby_doran}. We relate the comoving wavenumber, measured in Planck units, to the physical scale of the perturbation today via
        \begin{equation}
        k_{\mathrm{today}} = \frac{k}{a_0},
        \end{equation}
       where~$a_0$ is the present day scale factor, given in terms of the present day reduced Hubble parameter,~$h \equiv H_{0} /\left(100 \mathrm{km}\, \mathrm{s}^{-1} \mathrm{Mpc}^{-1}\right)$, and the present day density in curvature,~$\Omega_{k,0}$, by
       \begin{equation}
        a_0 \approx 4.3\times 10^4 \Big(\frac{h}{0.7} \Big)^{-1} \Big| \frac{\Omega_{k,0}}{0.01}\Big|^{-\frac{1}{2}} \mathrm{Mpc}. 
       \end{equation}
       
\begin{figure*}
            \centering
            \includegraphics{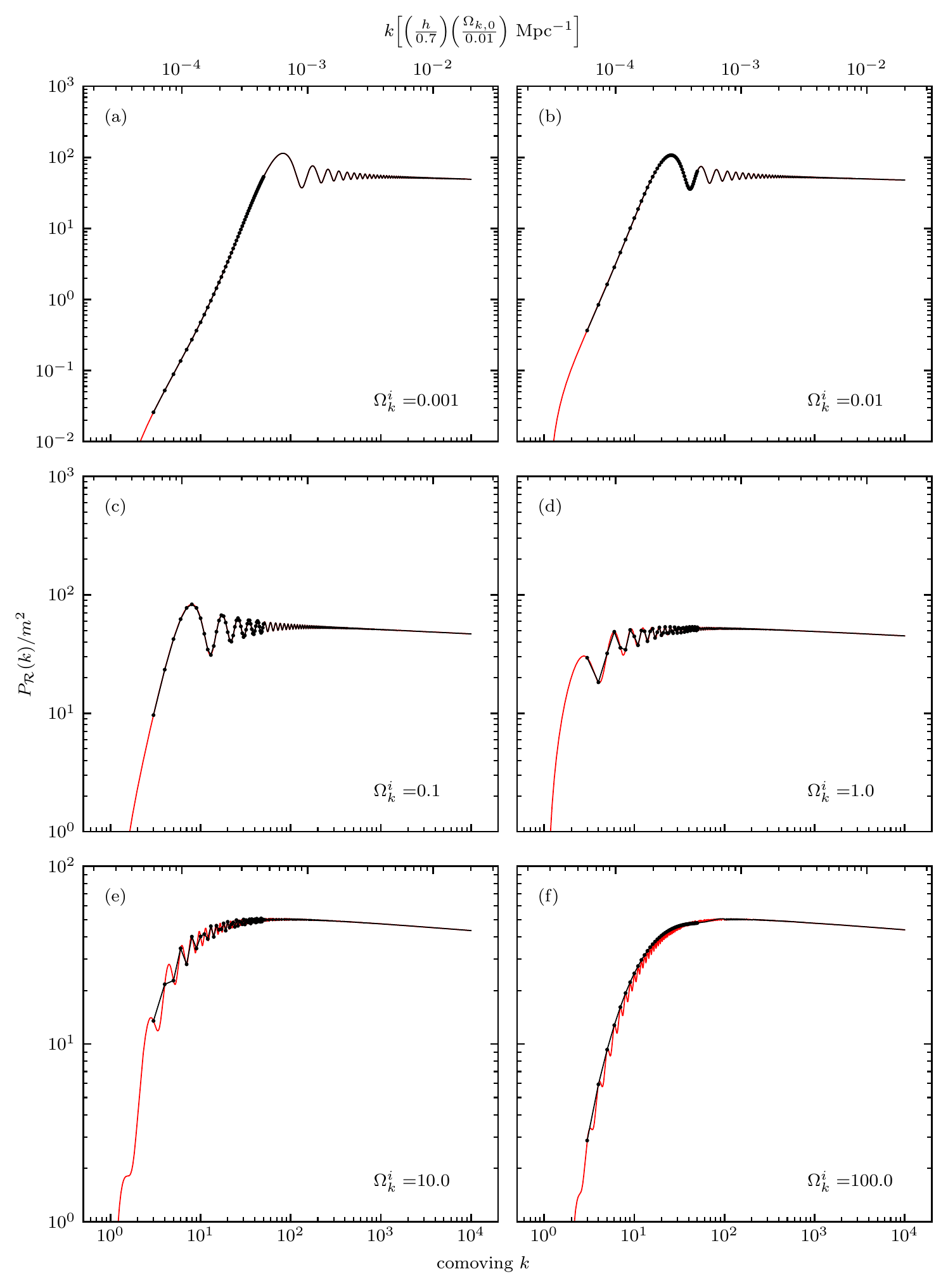}
            \caption{Scalar primordial power spectra in universes with varying initial curvature. The start of inflation,~$N_i$ is adjusted to vary with the curvature at the start of inflation,~$\Omega_k^i$, such that the total e-folds of inflation,~$N_{\mathrm{tot}}=60$ is constant. In curved universes, only integer values of comoving~$k$ are allowed, with~$k \geq 3$ (continuous line with~$k \leq  50$ highlighted), but for clarity we include the continuous spectrum (dashed line). The modes are started from the Bunch--Davies vacuum.  }
            \label{fig:kd-pps-closed}
        \end{figure*}

       Figure \ref{fig:kd-pps-closed} shows the resulting primordial power spectra for various values of initial curvature~$\Omega_k^i$, each with an associated spectrum treating comoving~$k$ as a continuous variable plotted underneath. Calculating the spectra with the RKWKB method provided roughly three orders of magnitude reduction in computing time compared to Runge--Kutta-like methods. 

\section{Limitations}\label{sec:limitations}
    
    When applying our solver to a problem, it is worth considering whether the solver's performance would be limited by strict accuracy requirements or by a non-ideal choice of independent or dependent variable.

    As shown in Figure \ref{fig:burst_runtime_rtol}, the algorithm's runtime scales up gently in the relative tolerance range~$[ 10^{-4}, 10^{-6}]$. The user is therefore advised to use the solver if such accuracies are acceptable for the problem in question. If the problem requires~$\mathrm{rtol} < 10^{-6}$, one would need a higher-order Runge--Kutta pair as an alternative solver to WKB, such as a (7,8) pair used by the \texttt{NAG} Library.
    
    As mentioned in Section \ref{sec:airy}, the solver will not be able to fulfil the accuracy requirements if at any time the integral(s)~$[S_i]_{t}^{t+h}$ exceed~$\sim $~$10^{12}$.
    Care needs to be taken especially with the first term in the WKB series,~$ \pm i\int_{t}^{t+h} \omega dt$, as this is expected to be the largest in a region where the WKB approximation is appropriate.
    The reason underlying this limit is that the solver needs to compute the exponential of this large imaginary term, which requires large accuracy modulo~$2\pi$.
    Storing such large numbers accurately is limited by machine (double) precision, and the solver might start accumulating error. In Section \ref{sec:airy} the stepsize and frequency become so large at~$t > 10^8$ that this limit is reached. 
   
    Finally, the solver is only efficient if in some region the frequency is slowly varying, therefore care needs to be taken to choose an appropriate independent-dependent variable pair if the problem allows. In our cosmological examples~$\mathcal{R}_k$ was chosen for its freezing-out property which makes the computation easy outside of the horizon, and~$N$ instead of cosmic time~$t$ because it does not span several orders of magnitude during the integration and gives a remarkably simple~$\ln \omega$. The frequency and friction term in terms of~$N$ are smooth and slowly varying, which allowed for them to be well-approximated by linear interpolation on a sufficiently fine, evenly spaced grid.
    
    The most immediate future generalisation of the algorithm would involve extending it to several dimensions, so that one could solve a coupled set of oscillatory differential equations. However, exploratory investigation revealed this task to be more difficult than anticipated~\cite{jamie_project}.
    
    A reduction in runtime and simplification of the stepping procedure would be possible if at small stepsizes, the step proposed by the algorithm using the WKB approximation reduced to a Runge--Kutta step of similar order. At present, the small-stepsize limit of the WKB steps is Euler's method. Euler's method not being efficient enough for practical use makes it necessary to take an alternative higher order Runge--Kutta step, which adds computational overhead.
    
    It is worth noting that due to the oscillatory nature of equations, \texttt{oscode} only obtains the solution at the start and end of integration ($t_{\mathrm{start}}$,~$t_{\mathrm{end}}$), and at a set of intermediate points determined by the solver.
    If the solution is required at a set of specific points, then it can be acquired by running the solver multiple times with~$t_{\mathrm{start}}$ and~$t_{\mathrm{end}}$ coinciding with the desired set. We are considering how the solution could be obtained at any given point within~$[t_{\mathrm{start}},t_{\mathrm{end}}]$ from the solutions (\ref{eq:gen-step-x}) and (\ref{eq:gen-step-dx}), and plan to include this feature in a future release.
        
\section{Conclusions}\label{sec:conclusion}

    We have presented a novel numerical solver for second-order, ordinary differential equations that can be written in the form of a one-dimensional oscillator, with a time-varying frequency and friction term that do not necessarily have a closed form.
    We have shown that the solver is significantly more efficient than other known methods if the frequency varies slowly over some part of the integration range, even if it is extremely large, because the solver can exploit the WKB approximation in these cases to traverse many oscillations at once.
    We have also shown that the solver can detect regions where the WKB approximation is not valid, and can dynamically switch to a Runge--Kutta integrator. 
    
    We demonstrated the above properties on several examples, the Airy equation, a more complex `burst' equation, and the Schr\"odinger equation, for which the frequency term can be written as a function of time, and the Mukhanov--Sasaki equation where both the frequency and the friction term need to be computed numerically in advance.
    In the case of the Mukhanov--Sasaki equation, we compared the solver's performance to that of \texttt{BINGO}, a highly efficient Fortran code that computes the scalar bi-spectrum by first computing a primordial power spectrum of scalar curvature perturbations using a fast Runge--Kutta solver available from \texttt{RKSUITE}. 
    We measured for each wavenumber~$k$ how long each code takes to compute a solution to the Mukhanov--Sasaki equation from sub-horizon ($k/aH = 100$) to super-horizon ($k/aH = 0.01$) times with all parameters identical, and found that our solver takes constant time in~$k$, being approximately twice as fast as \texttt{BINGO} in the observational range.
    If integration started when modes were deeper inside the Hubble horizon, the performance difference increases dramatically. 
    To prove this, we demonstrated that our solver is capable of integrating each mode from a single fixed time through horizon entry and exit, starting from kinetically dominated initial conditions for both the smooth, isotropic universe and the perturbations. 
    We further computed primordial power spectra for closed universes with varying initial curvature, a family of models in which the oscillatory equation of motion cannot be transformed into a first-derivative-free form, making it impossible to be solved with the efficient non-Runge--Kutta method developed in~\cite{Haddadin}.
    
\section*{Acknowledgements}
    
    FJA thanks Lukas Hergt for his suggestions about the algorithm and the numerous discussions on it. She also thanks STFC for their support. WJH thanks Gonville \& Caius college for their continuing support via a college research fellowship.

\bibliographystyle{elsarticle-num} 
\bibliography{references.bib}

\appendix
        
\section{Estimating the error in RK and WKB steps}
\label{sec:error-estimates}

The RK and WKB steps each give~$x$ and~$\dot{x}$ (referred to by their subscripts), and the difference between the~$4^{\mathrm{th}}$ and~$5^{\mathrm{th}}$ order RK steps gives an error on them. Estimating the error on a WKB step is less straightforward, and we decided to use the larger of two error estimates which dominate in different limits, as discussed below.

The obvious equivalent error estimate on WKB steps,~$\Delta x_{\mathrm{WKB}}$ and~$\Delta \dot{x}_{\mathrm{WKB}}$, is the difference between an~$N^{\mathrm{th}}$ and~$(N-1)^{\mathrm{th}}$ order estimate, where~$N$ refers to the highest-order~$S$-term in (\ref{eq:s-solutions}) included in the WKB expansion.
This estimate is a good proxy for the validity of the WKB approximation because it can signal the breakdown of the relations (\ref{eq:asymp-rels}), but in a region where they hold, it is expected that the numerical error in the~$S_i(t)$ will dominate~$\Delta x$ and~$\Delta \dot{x}$. We therefore estimate the error on the WKB step arising from the imperfect numerical integration of~$\dot{S}_i(t)$ as
\begin{align}\label{eq:err-wkb-x}
\Delta x_{\mathrm{WKB}} &= A_{+}\Delta f_{+} + A_{-}\Delta f_{-},\\
\Delta f_{\pm} &= f_{\pm} \sum_{i=0}^{n} \Delta [S_i]^{t+h}_t,
\end{align}
and
\begin{align}\label{eq:err-wkb-dx}
\Delta \dot{x}_{\mathrm{WKB}} &= B_{+}\Delta \dot{f}_{+} + B_{-}\Delta \dot{f}_{-},\\ 
\Delta \dot{f}_{\pm} &= \Delta f_{\pm}\frac{\dot{f}_{\pm}}{f_{\pm}}.
\end{align}
Note that in the above,~$f$ and its derivatives are evaluated at~$t+h$ according to (\ref{eq:gen-step-a})--(\ref{eq:gen-step-b}), and that it is assumed that the numerical integration of~$\dot{S}_i$ are the \emph{only} sources of error, i.e.\ the~$\dot{S}_i$ can be acquired perfectly.

\section{Runge--Kutta methods with Gauss--Lobatto stencils}
\label{sec:rk-tableau}

In this section we present the Butcher tableau of the two Runge--Kutta methods used in the solver. Table \ref{table:rk-glo-4} contains the coefficients of the 4-stage, 4${}^{\mathrm{th}}$ order method, and Table \ref{table:rk-glo-5} contains those of the  6-stage, 5${}^{\mathrm{th}}$ order one.

    \begin{table}[h]
        \centering
        \tiny{
        \begin{tabular}{l|cccc}
        0 & & & & \\
        $ \frac{1}{2}\left(1-\sqrt{\frac{3}{7}}\right) $ & $
        \frac{1}{2}\left(1-\sqrt{\frac{3}{7}}\right)  $ & & & \\
        $ \frac{1}{2}\left(1 + \sqrt{\frac{3}{7}}\right) $ & $
        -\frac{1}{4}\left(3+5\sqrt{\frac{3}{7}}\right)  $ & $ \frac{1}{4}\left( 5 + \sqrt{21} \right)  $ & &\\
        $ 1 $ & $ -\frac{1}{4}\left( 3 + 7\sqrt{21}\right) $ & $ -\frac{1}{4}\left( 21 + 5\sqrt{21} \right)
         $ & $ \frac{1}{14}\left( -1 + \sqrt{\frac{3}{7}}\right)  $ & \\ \hline
        & $ -\frac{1}{12}  $ & $ \frac{7}{12}  $ & $
        -\frac{1}{12} $ & $ -\frac{1}{12}  $
        \end{tabular}
        \caption{Butcher tableau for the 4-stage,~$4^{\mathrm{th}}$ order Runge--Kutta method used in the solver, based on 4 out of 5 stencil points of a Gauss--Lobatto quadrature with~$n=5$ stencil points.}
        \label{table:rk-glo-4}
        }
    \end{table}
   
    \begin{table}[h]
        \centering
        \footnotesize{
        \begin{tabular}{l|c}
        $c_1$ & 0  \\
        $c_2$ & $\frac{1}{2}\left(1-\sqrt{\frac{1}{3} + \frac{2\sqrt{7}}{21}}\right)$ \\
        $c_3$ & $\frac{1}{2}\left(1-\sqrt{\frac{1}{3} - \frac{2\sqrt{7}}{21}}\right)$ \\
        $c_4$ & $ \frac{1}{2}\left(1+\sqrt{\frac{1}{3} - \frac{2\sqrt{7}}{21}}\right)$ \\
        $c_5$ & $ \frac{1}{2}\left(1+\sqrt{\frac{1}{3} + \frac{2\sqrt{7}}{21}}\right) $ \\
        $c_6$ & $1$ \\ 
        $a_{21}$ & $ 0.117472338035267$ \\
        $a_{31}$ & $ -0.186247980065150 $ \\
        $a_{32}$ & $ 0.543632221824827 $ \\
        $a_{41}$ & $ -0.606430388550828  $ \\
        $a_{42}$ & $ 1  $ \\
        $a_{43}$ & $ 0.249046146791150  $ \\
        $a_{51}$ & $ 2.89935654001573 $ \\
        $a_{52}$ & $  -4.36852561156624  $ \\
        $a_{53}$ & $  2.13380671478631 $ \\
        $a_{54}$ & $  0.217890018728924 $ \\
        $a_{61}$ & $ 18.6799634999572 $\\
        $a_{62}$ & $ -28.8505778397313  $ \\
        $a_{63}$ & $ 10.7205340842092  $ \\
        $a_{64}$ & $ 1.41474175650804 $\\
        $a_{65}$ & $-0.964661500943270$ \\
        
        $b_1$ & $ 0.112755722735172$ \\
        $b_2$ & $0$ \\
        $b_3$ & $0.506557973265535$ \\
        $b_4$ & $ 0.0483004037699511$ \\
        $b_5$ & $0.378474956297846$ \\
        $b_6$ & $-0.0460890560685063$\\
        \end{tabular}
        }
        \caption{Butcher tableau for the 6-stage, 5${}^{\mathrm{th}}$ order Runge--Kutta method used by the solver. The timepoints of evaluation are the 6 stencil points used for Gauss--Lobatto quadrature with~$n=6$. }
        \label{table:rk-glo-5}
    \end{table}    

\section{Stepping procedure}
\label{sec:stepping}
   
    Let us summarise the different error estimates:
    \begin{itemize}
    \item{$\Delta x_{\mathrm{RK}}$,~$\Delta \dot{x}_{\mathrm{RK}}$: error on RK step,}
    \item{$\Delta x_{\mathrm{WKB}}$,~$\Delta \dot{x}_{\mathrm{WKB}}$: error on WKB step from computing~$[S_i]^{t+h}_t$ numerically,}
    \item{$\Delta x^{\mathrm{t}}_{\mathrm{WKB}}$,~$\Delta \dot{x}^{\mathrm{t}}_{\mathrm{WKB}}$: error on WKB step} from truncation of WKB series.
    \end{itemize}
    In order for the solver to switch successfully to the most suitable method dynamically, and adapt the stepsize to stay within the error bound required, it has to determine two things:
    \begin{enumerate}
    \item Which step (RK or WKB) to choose that yields the largest possible next stepsize within acceptable tolerance?
    \item What should the size of the next step be?
    \end{enumerate}
    
    The answer to 1. requires forecasting the error progression of both methods with the stepsize, i.e.\ requires knowledge of~$\Delta x(h)$ and~$\Delta \dot{x}(h)$.
    For the RK step this behaviour is known to be a power-law, and for WKB steps we shall assume two separate power-laws with different exponents~$n_{\mathrm{WKB}}$ and~$n^t_{\mathrm{WKB}}$, for when the dominant error on WKB steps arises from the numerical integrals and the truncation of the asymptotic series, respectively.
    First, the dominant error on each type of step is determined,
    \begin{align}\label{eq:rk-wkb-delta1}
    \Delta_{\mathrm{RK}} = \mathrm{max}(&\epsilon, \Delta x_{\mathrm{RK}} , \Delta \dot{x}_{\mathrm{RK}}), \\ 
    \Delta_{\mathrm{WKB}} = \mathrm{max}(&\epsilon, \Delta x_{\mathrm{WKB}} , \Delta \dot{x}_{\mathrm{WKB}},\nonumber \\ & \Delta x^t_{\mathrm{WKB}}, \Delta \dot{x}^t_{\mathrm{WKB}}),
    \end{align}
    where~$\epsilon$ is a small number close to machine precision, for safety. The type of dominant error on the WKB step, `\emph{truncation}', or `\emph{integral}' is recorded.
    Starting from a current stepsize~$h$, the largest possible steps within the error bound~$tol$ are then 
    \begin{equation}\label{eq:switching} 
    \begin{split}
    &h_{\mathrm{RK}} = h\times \left( \frac{\mathrm{tol}}{\Delta_{\mathrm{RK}}}\right)^{1/n_{\mathrm{RK}}}, \\
    &h_{\mathrm{WKB}}=h \begin{cases}
    \left( \frac{\mathrm{tol}}{\Delta_{\mathrm{WKB}}}\right)^{1/n^t_{\mathrm{WKB}}}, \, \text{if \emph{`truncation'}},\\
    \left( \frac{\mathrm{tol}}{\Delta_{\mathrm{WKB}}}\right)^{1/n_{\mathrm{WKB}}} \, \mathrm{otherwise.}
    \end{cases}
    \end{split}
    \end{equation}
    The step with the larger stepsize will then be chosen as a trial step, but is not yet accepted. The next stepsize is then predicted. If the chosen method is RK, this next stepsize is simply 
    \begin{equation}\label{eq:hnext-rk}
    h_{\mathrm{next}} = h_{\mathrm{RK}}. 
    \end{equation}
    If the chosen method is WKB however (i.e.\ the truncated WKB series was deemed sufficient to approximate the solution), the error arising from truncation of the WKB series will be ignored:
    \begin{gather}
    \mathrm{redefine} \; \Delta_{\mathrm{WKB}} \; \mathrm{as} \nonumber \\
    \Delta_{\mathrm{WKB}} = \mathrm{max}(\epsilon, \Delta x_{\mathrm{WKB}} , \Delta \dot{x}_{\mathrm{WKB}}), \label{eq:hnext-wkb-1}\\
    \mathrm{then} \quad h_{\mathrm{next}} = h\left( \frac{\mathrm{tol}}{\Delta_{\mathrm{WKB}}}\right)^{1/n_{\mathrm{WKB}}}. \label{eq:hnext-wkb-2}
    \end{gather}
    
    \begin{figure}[h]
    \centering
    \includegraphics{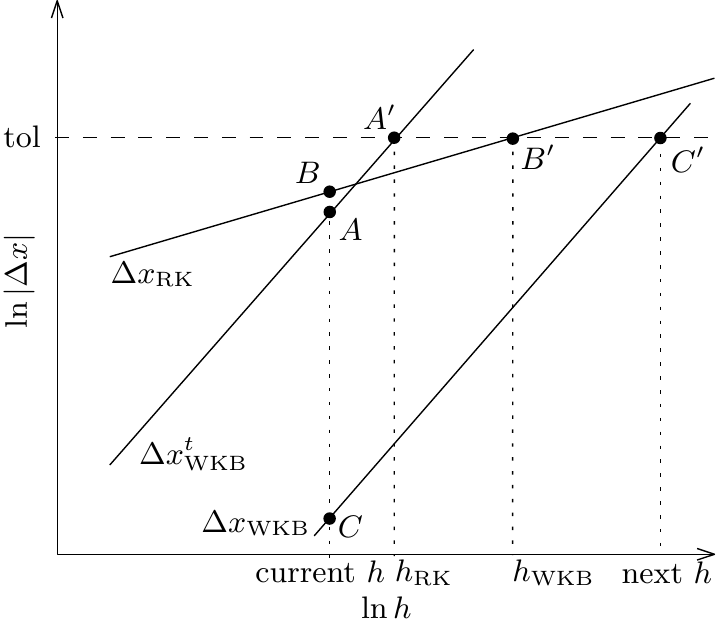}
    \caption{Schematic plot of the assumed error progression in RK and WKB steps with increasing stepsize~$h$. After the steps have been calculated from~$t$ to~$t+h_{\mathrm{current}}$, the errors of each method are shown by points~$A$,~$B$ and~$C$, the latter two arising from the truncation of the WKB asymptotic series and the numerical integrals present in the series, respectively. The dominant type of error on the WKB step in this case is the `\emph{truncation}'.
    Assuming power-law behaviour in the errors for both steps with different exponents for each type of error, the next largest stepsize within the required tolerance `tol' would be~$h_{\mathrm{RK}}$ and~$h_{\mathrm{WKB}}$, marked by points~$A'$ and~$B'$. Since~$h_{\mathrm{WKB}} > h_{\mathrm{RK}}$, the algorithm in this case chooses the WKB step. The size of the next step is therefore determined solely on the basis of the `\emph{integral}' error, marked by~$C$, and is going to be the projection of~$C'$. Since this next stepsize is larger than the previous, the step is accepted.} 
    \label{fig:wkb-rk-errs}
    \end{figure}
    
    This process is illustrated in Figure \ref{fig:wkb-rk-errs}.
    Finally, if~$h_{\mathrm{next}} > h$, the current error did not exceed the tolerance limit and the step is accepted. Otherwise the step is rejected and one must ensure that the step is re-attempted with sufficiently small~$h$. The new stepsize in both cases is calculated via
    \begin{align}\label{eq:hnext-unsucc}
    h_{\mathrm{next}} = h \times \begin{cases}
    \left( \frac{\mathrm{tol}}{\Delta_{\mathrm{RK}}}\right)^{1/(n_{\mathrm{RK}}-1)},\\
    \left( \frac{\mathrm{tol}}{\Delta_{\mathrm{WKB}}}\right)^{1/(n_{\mathrm{WKB}}-1)} \, \text{if `\emph{integral}', }\\
    \left( \frac{\mathrm{tol}}{\Delta_{\mathrm{WKB}}}\right)^{1/(n^t_{\mathrm{WKB}}-1)} \, \text{if `\emph{truncation}'.}
    \end{cases}
    \end{align} 
    This ensures~$h$ is decreased after rejected steps and increased following accepted ones.
    
    The slopes of the errors as functions of~$h$ in Figure \ref{fig:wkb-rk-errs} were not chosen at random.
    The error in a 6-stage, 5${}^{\mathrm{th}}$-order RK method goes as~$h^5$ for~$h < 1$ and~$h^6$ for~$h > 1$.
    The error arising from truncation of the WKB series, upon entering a region well-approximated by the WKB expansion, is expected to be proportional to~$h$. This can be understood starting from the relative error on~$x$ based on~\cite{BenderOrszag},
    \begin{equation}
    \frac{\Delta x}{x} \sim S_{N+1},
    \end{equation}
    for an~$N^{\mathrm{th}}$ order WKB estimate. For~$N \geq 1$,~$S_{N+1}$ is a numerical integral of a small and nearly constant quantity, and is therefore~${\propto \dot{S}_{N+1}h}$. The error on WKB steps arising from the integrals~$[S_i]_t^{t+h}$ are on the other hand expected to go roughly as the errors on the integrals themselves (see (\ref{eq:err-wkb-x})). Although more difficult to predict, this is expected to be dominated by the imperfect evaluations of the integrands, which contain numerical derivatives. The largest of these are the first derivatives, which will have an error~${\propto h^{n_s - 1}}$. Since the algorithm uses the~$n=6$ Gauss--Lobatto evaluations to calculate all derivatives, we set~$n_s = 6$.
    
    By the above reasoning, the algorithm by default has 
    \begin{equation}\label{eq:defaults}
    n_{\mathrm{RK}} = 5, \quad n_{\mathrm{WKB}}=5, \quad n^t_{\mathrm{WKB}}=2,
    \end{equation}
    but the user can set these parameters to better fit the problem in question.
    For example, for optimal step acceptance/rejection ratio, for all burst examples in Section \ref{sec:burst} we set~$n_{\mathrm{WKB}}=8$,~$n^t_{\mathrm{WKB}}=1$.

\section{Summary}
\label{sec:summary}

The algorithm goes through the following steps:
\begin{enumerate}
    \item Stepping from~$t$ to~$t+h$, evaluate~$\omega$ and~$\gamma$ at the Gauss--Lobatto stencil points for~$n=6$ and~$n=5$, a total of 9 different points.
    \item Use the Butcher tableaux \ref{table:rk-glo-5} and \ref{table:rk-glo-4} to construct a RK step in~$x$ and~$\dot{x}$, and use the difference as the error~$\Delta x_{\mathrm{RK}}$.
    \item Use finite difference methods to evaluate all necessary derivatives of~$\omega$ and~$\gamma$ needed for the derivatives of terms ($\dot{S}_i$) in the WKB series (\ref{eq:s-solutions}). 
    \item Use Gauss--Lobatto quadrature with~$n=6$ and~$n=5$ to evaluate the terms in the WKB series and their errors, taken as the difference.
    \item Construct an~$N^{\mathrm{th}}$ and~$(N-1)^{\mathrm{th}}$ order WKB step in~$x$ and~$\dot{x}$.
    \item Compute the `\emph{truncation}' error on WKB steps as the difference between the~$N^{\mathrm{th}}$ and~$(N-1)^{\mathrm{th}}$ order estimates, and the `\emph{integral}' from (\ref{eq:err-wkb-x}) and (\ref{eq:err-wkb-dx}).
    \item Find the dominant error and its type based on (\ref{eq:rk-wkb-delta1}), and choose between RK/WKB methods based on (\ref{eq:switching}). 
    \item Predict the next stepsize,~$h_{\mathrm{next}}$, based on (\ref{eq:hnext-rk})--(\ref{eq:hnext-wkb-2}).
    \item \textbf{If}~$h_{\mathrm{next}} > h$, accept the step, and update~$x$,~$\dot{x}$, and~$t$.
    \item \textbf{Otherwise}, reject the step and calculate~$h_{\mathrm{next}}$ according to (\ref{eq:hnext-unsucc}). 
    \item Update~$h$ to~$h_{\mathrm{next}}$.
    \item Repeat steps 1--11.
\end{enumerate}
  
\end{document}